\documentclass[runningheads]{llncs}
\usepackage{amsmath}
\usepackage{mathtools}
\usepackage{amssymb}
\usepackage{amsfonts}
\usepackage{amssymb}
\usepackage[framemethod=tikz]{mdframed}
\usepackage{graphicx}
\usepackage{stmaryrd}
\usepackage{semantic}
\usepackage{color}
\usepackage{subcaption}
\usepackage[colorlinks=true,linkcolor=blue,pdftex,bookmarks=true,bookmarksopen,bookmarksdepth=2]{hyperref}
\usepackage{bookmark}
\usepackage{marginnote}
\usepackage{url}
\usepackage{marvosym}
\usepackage{pgf}
\usepackage{pgf}
\usepackage{tikz}
\usepackage{scalerel}
\usetikzlibrary{arrows,automata,shapes,backgrounds}

\usepackage[most]{tcolorbox}

\tcbset{
	frame code={}
	center title,
	left=0pt,
	right=0pt,
	top=0pt,
	bottom=0pt,
	colback=gray!30,
	colframe=white,
	width=\dimexpr\textwidth\relax,
	enlarge left by=0mm,
	boxsep=5pt,
	arc=0pt,outer arc=0pt,
}

\graphicspath{ {fig/} }
\author{Rocco De Nicola\inst{1}, Alessandro Maggi\inst{1} \and Joseph Sifakis\inst{2}}
\institute{IMT School for Advanced Studies Lucca \and Universit\'e Grenoble Alpes}
\title{DReAM: Dynamic Reconfigurable Architecture Modeling (full paper)}

\DeclareRobustCommand{\LANG}{\textsf{DReAM}} %Dynamic Reconfigurable Architecture Modeling
\DeclareRobustCommand{\PIL}{\textsf{PIL}}
\DeclareRobustCommand{\PILOPS}{\textsf{PILOps}}

\DeclareRobustCommand{\VALUES}{\mathsf{V}}

\definecolor{grey}{RGB}{180,180,180}

\usetikzlibrary{calc,decorations.pathmorphing,shapes}
\newcounter{sarrow}
\newcommand\xrsquigarrow[2][10]{%
	\stepcounter{sarrow}%
	\mathrel{\begin{tikzpicture}[baseline= {( $ (current bounding box.south) + (0,-0.5ex) $ )}]
		\node[inner sep=.5ex] (\thesarrow) {$\scriptstyle #2$};
		\path[draw,<-,decorate,
		decoration={zigzag,amplitude=0.8pt,segment length=1.2mm,pre=lineto,pre length=#1pt, post=lineto, post length=#1pt}] 
		(\thesarrow.south east) -- (\thesarrow.south west);
		\end{tikzpicture}}%
}

\DeclareMathOperator*{\bigand}{\scalerel*{\&}{\textstyle\sum}}

\DeclareRobustCommand{\RULE}[2]{#1 \rightarrow #2}
\DeclareRobustCommand{\evRULE}[2]{#1 \big\lbrace #2 \big\rbrace}
\DeclareRobustCommand{\irRULE}[4]{#1 \big[ #2 , #3 , #4 \big]}

\DeclareRobustCommand{\irClRULE}[4]{#1 \big\lbrace \big[ #2 , #3 , #4 \big] \big\rbrace}
\DeclareRobustCommand{\rcClRULE}[3]{#1 \big\lbrace \RULE{#2}{#3} \big\rbrace}
\DeclareRobustCommand{\AtMost}[2]{AtMost(#1,#2)}
\DeclareRobustCommand{\MOTIFconf}[6][M]{\left\langle {#2}_{#1}.s{#3}, {#2}_{#1}.\sigma{#4}, Map{#5}_{#1}, \MVAt{#6}_{#1} \right\rangle}
\DeclareRobustCommand{\EXECOPS}[2][a,\Gamma]{\llbracket #2 \rrbracket_{#1}}
\DeclareRobustCommand{\MODELS}[3]{#1 \models_{#2} #3}
\DeclareRobustCommand{\NOTMODELS}[3]{#1 \nvDash_{#2} \ #3}
\DeclareRobustCommand{\CONCRETE}[2][\Gamma]{\left\langle #2 \right\rangle_{#1}}

\DeclareRobustCommand{\VarDef}[2]{#1 \ : \ \mathsf{#2}}

\DeclareRobustCommand{\IDLE}[1][]{idle_{#1}}
\DeclareRobustCommand{\TRUE}{\mathbf{true}}
\DeclareRobustCommand{\FALSE}{\mathbf{false}}
\DeclareRobustCommand{\NOT}[1]{\neg #1}

\DeclareRobustCommand{\@}[1]{\MVAt\left(#1\right)}

\DeclareRobustCommand{\ASSIGN}[3]{#1#2 := #3}

\DeclareRobustCommand{\CREATE}[2]{create\left(#1,#2\right)}
\DeclareRobustCommand{\DELETE}[1]{delete\left(#1\right)}
\DeclareRobustCommand{\ADDNODE}[1]{add\left(#1\right)}
\DeclareRobustCommand{\REMOVENODE}[1]{remove\left(#1\right)}
\DeclareRobustCommand{\ADDEDGE}[2]{add\left(#1,#2\right)}
\DeclareRobustCommand{\REMOVEEDGE}[2]{remove\left(#1,#2\right)}
\DeclareRobustCommand{\MOVE}[2]{move\left(#1,#2\right)}

\DeclareRobustCommand{\MIGRATE}[3]{migrate\left(#1,#2,#3\right)}

\DeclareRobustCommand{\IFTHEN}[2]{\mathsf{IF} \ \big(#1\big) \ \mathsf{THEN} \ #2}
\DeclareRobustCommand{\IFTHENELSE}[3]{\IFTHEN{#1}{#2} \ \mathsf{ELSE} \ #3}

\begin{document}
	\maketitle
	
	\begin{abstract}
Modern systems evolve in unpredictable environments and have to continuously adapt their behavior to changing conditions.  The ``\LANG{}'' (Dynamic Reconfigurable Architecture Modeling) framework, has been designed for modeling reconfigurable dynamic systems. 
It provides a rule-based language, inspired from Interaction Logic, which is  expressive and easy to use encompassing all aspects of dynamicity including parametric multi-modal coordination with creation/deletion of components as well as mobility. 
Additionally, it allows the description of both endogenous/modular and exogenous/centralized coordination styles and sound transformations from one style to the other.
The \LANG{} framework is implemented in the form of a Java API bundled with an execution engine. 
It allows to develop runnable systems combining the expressiveness of the rule-based notation together with the flexibility 
of this widespread programming language. 
		
	\end{abstract}

	% !TEX root =../DreAM-article.tex
\section{Introduction}
\label{sec:intro}

The ever increasing complexity of modern software systems has changed the perspective of software designers who now have to consider new classes of systems, consisting of a large number of interacting components and featuring complex interaction mechanisms. These systems are usually distributed, heterogeneous, decentralised and interdependent, and are operating in an unpredictable environments. They need to continuously adapt to changing internal or external conditions in order to efficiently use of resources and to provide adequate functionality when the external environment changes dynamically. Dynamism, indeed, plays a crucial role in these modern systems and it can be captured as the interplay of changes relative to the three features below: 
\begin{enumerate}
	\item the parametric description of interactions between instances of components for a given system configuration; 
	\item the reconfiguration involving creation/deletion of components and management of their interaction according to a given architectural style; 
	\item the migration of components between predefined architectural styles. 
\end{enumerate}

Architecture modeling languages should be equipped with concepts and mechanisms which are expressive and easy to use relatively to each of these features.

The first feature implies the ability of describing the coordination of systems that are parametric with respect to the numbers of instances of types of components; examples of such systems are Producer-Consumer systems with $m$ producers and $n$ consumers or Ring systems consisting of $n$ identical interconnected components.

The second feature is related to the ability of reconfiguring systems by adding or deleting components and managing their interactions taking into account the dynamically changing conditions.  In the case of a reconfigurable ring this would require having the possibility of removing a component which self-detects a failure and of adding it back after recovery. Added components are subject to specific interaction rules according to their type and their position in the system. 
This is especially true for mobile components which are subject to dynamic interaction rules depending on the state of their neighborhood.

The third aspect is related to the vision of ``fluid architectures'' \cite{garlan2014software} or ``fluid software'' \cite{taivalsaari2014liquid} and builds on
the concept that applications and objects live in an environment (we call it a \emph{motif}) corresponding to an architectural style that is characterized by specific coordination and reconfiguration rules. Dynamicity of systems is modelled by allowing applications and objects to migrate among motifs and such dynamic migration allows a disciplined, easy-to-implement, management of dynamically changing coordination rules. 
For instance, self-organizing systems may adopt different coordination motifs to adapt their behavior and guarantee global properties.

The different approaches to architectural modeling and the new trends and needs are  reviewed in detailed surveys such as \cite{bradbury2004organizing,oreizy1998issues,malavolta2013industry,butting2017classification,medvidovic2007moving}. 
Here, we consider two criteria for the classification of existing approaches: \emph{exogenous vs. endogenous} and \emph{declarative vs. imperative} modeling.

\emph{Exogenous modeling} considers that components are architecture-agnostic and respect a strict separation between a component behavior and its coordination. 
Coordination is specified globally by coordination rules applied to sets of components. 
The rules involve synchronization of events between components and associated data transfer. 
This approach is adopted by Architecture Description Languages (ADL) \cite{malavolta2013industry}.
It has the advantage of providing a global view of the coordination mechanisms and their properties.

\emph{Endogenous modeling} requires adding explicit coordination primitives in the code describing components' behavior. Components are composed through their interfaces, which expose their coordination capabilities. 
An advantage of endogenous coordination is that it does not require programmers to explicitly build a global coordination model. 
However, validating a coordination mechanism and studying its properties becomes much harder without such a model.

\emph{Conjunctive modeling} uses logics to express coordination constraints between components. 
It allows in particular modular description as one can associate with each component its coordination constraints. 
The global system coordination can be obtained in that case as the conjunction of individual constraints of its constituent components. 

\emph{Disjunctive modeling} consists in explicitly specifying system coordination as the union of the executable coordination mechanisms such as semaphores, function call and connectors.

Merits and limitations of the two approaches are well understood.
Conjunctive modeling allows abstraction and modular description but it involves the risk of inconsistency in case there is no architecture satisfying the specification.

\begin{figure}[h]
	\centering
	\includegraphics[width=0.9\linewidth]{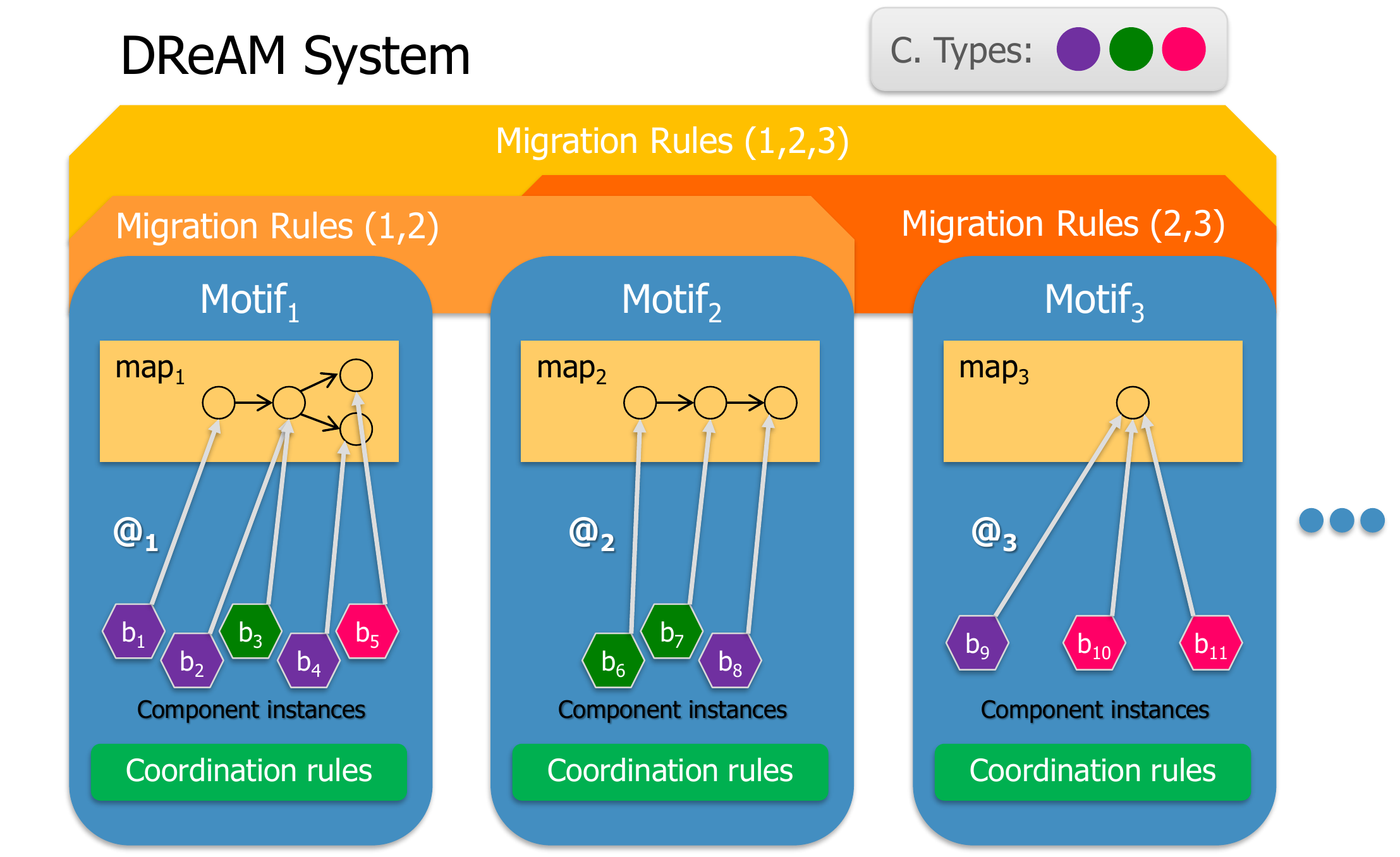}
	\caption{Overview of a \LANG{} system}
	\label{fig:dream-overview}
\end{figure}

This paper introduces the \LANG{} framework for modeling Dynamic Reconfigurable Architectures. 
\LANG{} uses a logic-based modeling language that encompasses the four styles mentioned above as well as the three mentioned features. 
A system consists of instances of types of components organized in a collection of motifs. 
Component instances can migrate between motifs depending on global system conditions. 
Thus, a given type of component can be subject to different rules when it is in a ``ring'' motif or in a ``pipeline'' one. 
Using motifs allows natural description of self-organizing systems (see Fig.\ref{fig:dream-overview}).

Coordination rules in a motif involve an interaction part and an associated operation. 
The former is modeled as a formula of the first order Interaction Logic \cite{bliudze2008algebra} used to specify parametric interactions between instances of types of components. 
The latter specifies transfer of data between the components involved in the interaction. 
In this way, we can characterize parametric coordination between classes of components. 
The rules allow both conjunctive and disjunctive specification styles.
We study to what extent a mathematical correspondence can be established between the two styles. 
In particular, we will see that conjunctive specifications can be translated into equivalent disjunctive global specifications while the converse is not true in general.

To enhance expressiveness of the different kinds of dynamism, each motif is equipped with a map, which is a graph defining the topology of the interactions in this motif.
To parametrize coordination rules for the nodes of the map, an address function $\MVAt$ is provided defining the position $\MVAt(c)$ in the map of any component instance $c$ associated with the motif. 
Maps are also very useful to express mobility of components, in which case the connectivity relation of the map represents possible moves of components. 
Finally the language allows the modification of maps by adding or removing nodes and edges, as well as the dynamic creation and deletion of component instances.  

The paper is organized as follows. 

Section~\ref{sec:pil-framework} presents the Propositional Interaction Logic (\PIL{}) and its use to model static architectures when the involved components are transition systems. 
It studies the relationship between conjunctive and disjunctive style and shows that for each conjunctive model there exists an equivalent disjunctive model and conversely.
 
Section~\ref{sec:pilops-framework} lifts the results of the previous section to components and interactions with data. 
Coordination constraints are expressed in the \PILOPS{} language whose terms are guarded commands where guards are \PIL{} formulas and commands are operations on data. 
\PILOPS{} is the core language of the \LANG{} framework.
  
Section~\ref{sec:full-framework} provides a formal definition of the \LANG{} framework. 
Coordination constraints are expressed in a first order extension of \PILOPS{} which allows quantification over component variables involved in rules and guards. 
We define operational semantics for \LANG{} models and propose an abstract syntax for a domain-specific language encompassing the basic modeling concepts. 
We also describe the Java-based modeling and execution framework under development and provide illustrating examples and benchmarks.

Section~\ref{sec:relatedwork} discusses related work with a comparison between main representatives of existing frameworks.

The conclusion summarizes the main results and discusses avenues for their further extension and application to real-life dynamic systems with focus on autonomous and self-modifying systems.

	\section{Static architectures - the \PIL{} coordination language}
\label{sec:pil-framework}

We introduce the Propositional Interaction Logic (\PIL{}) \cite{bliudze2008algebra} used to model interactions between a given set of components.

\subsection{Components}
\label{subsec:pil-components}

A system model is the composition of interacting components which are labelled transition systems, where the labels are port names and the states are control locations. 
Components are completely coordination-agnostic, as there is no additional characterization to ports and control locations
beyond their names (e.g. we do not distinguish between input/output ports or synchronous/asynchronous components).

\begin{definition}[Component]\label{def:pil-component}
	Let $\mathcal{P}$ and $\mathcal{S}$ respectively be the domain of ports and control locations.
	A component is a transition system $B = (S, P, T)$ with	
	\begin{itemize}
		\item $S \subseteq \mathcal{S}$: finite set of control locations;
		\item $P \subseteq \mathcal{P}$: finite set of ports;
		\item $T \subseteq S \times P \cup \left\lbrace \IDLE \right\rbrace \times S$: finite set of transitions.
		Transitions $(s,p,s')$ are also denoted by $s \xrightarrow{p} s'$; $p \in P$ is the port offered for interaction, and each transition is labelled by a different port.
	\end{itemize}
	A component has a special port $\IDLE \notin P$ that is associated to implicit loop transitions $\lbrace s \xrightarrow{idle} s \rbrace_{s \in S}$. 
	This choice is made to simplify the theoretical development of our framework.
	Furthermore it is assumed that the sets of ports and control locations of different components are disjoint.
\end{definition}
A system definition is characterized by a set of components $B_i = \left( S_i,P_i,T_i \right)$ for $i \in \left[1,n\right]$.
The \emph{configuration} $\Gamma$ of a system is the set of the current control locations of each constituent component:
\begin{align}
\Gamma = \left\lbrace s_i \in S_i \right\rbrace_{i \in \left[1..n\right]}
\end{align}

Given the set of ports $\mathcal{P}$, an interaction $a$ is any finite subset of $\mathcal{P}$ such that no two ports belong to the same component.
The set of all interactions is isomorphic to $I(\mathcal{P})=2^{\mathcal{P}}$.

Given a set of components $B_1 \dots B_n$ and the set of interactions $\gamma$, we can define a system $\gamma \left(B_1, \dots, B_n\right)$ using the following operational semantics rule:
\begin{align}\label{eq:pil-op-semantics-1}
\inference{
	a \in \gamma \qquad
	\forall p \in a: s_i \xrightarrow{p} s'_i
}{
	\left\lbrace s_i \right\rbrace_{\left[1..n\right]} \xrightarrow{a} \left\lbrace s'_i \right\rbrace_{\left[1..n\right]}
}[]
\end{align}
where $s_i$ is the current control location of component $B_i$, and $a$ is an interaction containing exactly one port for each component $B_i$\footnote{Components $B_j$ not ``actively'' involved in the interaction will participate with their $\IDLE$ port s.t. $s'_j = s_j$.}.

\subsection{Propositional Interaction Logic (\PIL{})}

Let $\mathcal{P}$ and $\mathcal{S}$ be respectively the domains of ports and control locations.
The formulas of Propositional Interaction Logic $\PIL(\mathcal{P},\mathcal{S})$ are defined by the syntax:
\begin{align}\label{eq:pil-syntax}
\text{(\PIL{} formula)} \quad & \varPsi ::= p \in \mathcal{P} \: | \: \pi \: | \: \NOT{\varPsi} \:|\: \varPsi_1 \wedge \varPsi_2 
\end{align}
where $\pi : 2^\Gamma \mapsto \left\lbrace \TRUE,\FALSE \right\rbrace$ is a \emph{state predicate}.
We use logical connectives $\vee$ and $\Rightarrow$ with the usual meaning.

The models of the logic are interactions on $\mathcal{P}$ for a configuration $\Gamma$.
The semantics is defined by the following satisfaction relation $\MODELS{}{\Gamma}{}$ between an interaction $a$ and formulas:
\begin{alignat}{2}\label{eq:pil-sat-relation}
\MODELS{a}{\Gamma}{& \TRUE} &\quad&\text{for any } a \nonumber \\
\MODELS{a}{\Gamma}{& p} &&\text{if } p \in a \nonumber \\
\MODELS{a}{\Gamma}{& \pi} &&\text{if } \pi(\Gamma)=\TRUE \nonumber \\
\MODELS{a}{\Gamma}{& \varPsi_1 \wedge \varPsi_2} &&\text{if } \MODELS{a}{\Gamma}{\varPsi_1} \text{ and } \MODELS{a}{\Gamma}{\varPsi_2} \nonumber \\
\MODELS{a}{\Gamma}{& \NOT{\varPsi}} &&\text{if } \NOTMODELS{a}{\Gamma}{\varPsi}
\end{alignat}

A monomial $\bigwedge_{p\in I}p \wedge \bigwedge_{p\in J} \NOT{p}, I \cap J = \emptyset$ characterizes a set of interactions $a$ such that:
\begin{enumerate}
	\item the positive terms correspond to required ports for the interaction to occur; 
	\item the negative terms correspond to inhibited ports or the ports to which the interaction is ``closed''; 
	\item the non-occurring terms are optional ports.
\end{enumerate}
When the set of optional ports is empty, then the monomial is a single interaction and it is characterized by $\bigwedge_{p\in a}p \wedge \bigwedge_{p\notin a}\NOT{p}$.

Note that $\IDLE$ ports of components can appear in \PIL{} formulas.
Given a component with ports $P$ and idle port $\IDLE$, the formula $\IDLE \equiv \bigwedge_{p \in P} \neg p$, while $\neg \IDLE \equiv \bigvee_{p \in P} p$. 

As we can describe sets of interactions using \PIL{} formulas, we can redefine the operational semantics rule (\ref{eq:pil-op-semantics-1}) as follows:
\begin{align}\label{eq:pil-op-semantics-2}
\inference{
	\MODELS{a}{\Gamma}{\varPsi} \qquad
	\forall p \in a: s_i \xrightarrow{p} s'_i
}{
	\left\lbrace s_i \right\rbrace_{\left[1..n\right]} \xrightarrow{a} \left\lbrace s'_i \right\rbrace_{\left[1..n\right]}
}[]
\end{align}
where $\varPsi$ is a \PIL{} formula.

\subsection{Disjunctive vs. conjunctive specification style}

It is shown in \cite{bliudze2008algebra} how a function $\beta$ can be defined $\beta: I(P) \rightarrow PIL(P,S)$ associating with an interaction $a$ its characteristic \PIL{} formula $\beta(a)$.
For example, if $P=\left\lbrace p,q,r,s,t \right\rbrace$ then for the interaction $\left\lbrace p,q\right\rbrace$, $\beta(\left\lbrace p,q\right\rbrace) = p \wedge q \wedge \neg r \wedge \neg s \wedge \neg t$\footnote{For the sake of conciseness, from now on we will omit the conjunction operator on monomials.}.
For the set of interactions $\gamma$ caused by the broadcast of $p$ to ports $q$ and $r$, $\beta(\gamma) = p \neg s \neg t$.
For the set of interactions $\gamma$ consisting of the singleton interactions $p$ and $q$, $\beta(\gamma) = \left( p \neg q \vee \neg p q \right) \wedge \neg r \neg s \neg t$.
Finally $\beta(\left\lbrace\IDLE\right\rbrace) = \neg p \neg q \neg r \neg s \neg t$ as $\IDLE$ is the only port not belonging to $P$.

Note that the definition of the function $\beta$ requires knowledge of $P$.
This function can be naturally extended to sets of interactions $\gamma$: for $\gamma = \left\lbrace a_1, \dots , a_n \right\rbrace$, $\beta(\gamma) = \beta\left(a_1\right) \vee \dots \vee \beta\left(a_n\right)$.

A set of interactions is specified in \emph{disjunctive style} if it is described by a \PIL{} formula which is a disjunction of monomials.
A dual style of specification is the \emph{conjunctive style} where the interactions of a system are the conjunction of \PIL{} formulas.
A methodology for writing conjunctive specifications proposed in \cite{bliudze2008algebra} considers that each term of the conjunction is a formula of the form $p \Rightarrow \varPsi_p$, where the implication is interpreted as a causality relation: for $p$ to be true, it is necessary that the formula $\varPsi_p$ holds and this defines interaction patterns from other components in which the port $p$ needs to be involved.

For example, the interaction involving strong synchronization between $p_1$, $p_2$ and $p_3$ is defined by the formula $f_1 = \left(p_1 \Rightarrow p_2 \right) \wedge \left(p_2 \Rightarrow p_3 \right) \wedge \left(p_3 \Rightarrow p_1 \right)$.
Broadcast from a sending port $t$ towards receiving ports $r_1,r_2$ is defined by the formula $f_2 = \left(\TRUE \Rightarrow t \right) \wedge \left(r_1 \Rightarrow t \right) \wedge \left(r_2 \Rightarrow t \right)$.
The non-empty solutions are the interactions $t$, $tr_1$, $tr_2$ and $tr_1r_2$.

Note that by applying this methodology we can associate to a component with set of ports $P$ a constraint $\bigwedge_{p \in P} \left( p \Rightarrow \varPsi_p \right)$ that characterizes the set of interactions where some port of the component may be involved. 
So if a system consists of components $C_1, \dots,C_n$ with sets of ports $P_1,\dots, P_n$ respectively, then the \PIL{} formula $\bigwedge_{i \in \left[1,n\right]} \bigwedge_{p \in P_i} \left( p \Rightarrow \varPsi_p \right)$ expresses a global interaction constraint. 
Such a constraint can be put in disjunctive form whose monomials characterize global interactions. 
Notice that the disjunctive form obtained in that manner contains the monomial $\bigwedge_{p \in P} \neg p$, where $P = \bigcup_{i \in \left[1..n\right]}P_i$, which is satisfied by the interaction where every component performs the $\IDLE$ action. 
This trivial remark says that in the \PIL{} framework it is possible to express for each component separately its interaction constraints and compose them conjunctively to get global disjunctive constraints.

\begin{figure}[h]
	\centering
	\includegraphics[width=0.9\linewidth]{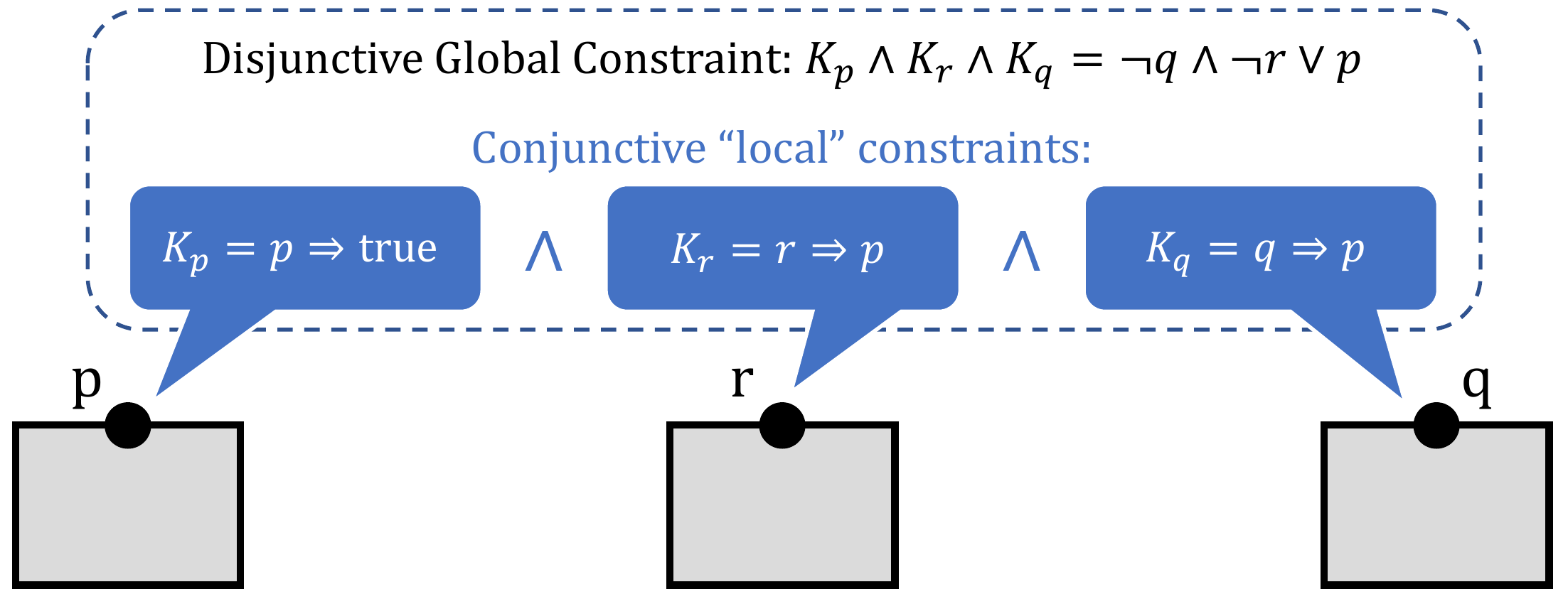}
	\caption{Broadcast example: disjunctive vs conjunctive specification}
	\label{fig:disjunctive-global-constraint}
\end{figure}

It is also possible to put in conjunctive style a disjunctive formula $\varPsi$ specifying the interactions of a system with set of ports $P$.
%For example, the formula $\bigwedge_{p \in P_i} p \vee \phi$ can be put in the form $\bigwedge_{p \in P}\left(p \Rightarrow \phi\right)$ and can be simplified to $\bigwedge_{p \in P}\left(p \Rightarrow \phi\left[p = \TRUE\right]\right)$.
%So, $\phi_p = \phi\left[p = \TRUE\right]$ obtained from $\phi$ by substituting $\TRUE$ to $p$.
%The conjunctive translation of $\phi$ will have a form $\bigwedge_{p \in P}\left(p \Rightarrow \phi_p \right) \wedge \left( \bigwedge_{p\in P} \neg p  \Rightarrow \FALSE \right)$, where the second term accounts for the fact that the disjunctive style prevents idling by default and in the first term $\phi_p = \phi\left[p = \TRUE\right]$ is obtained from $\phi$ by substituting $\TRUE$ to $p$. 
To translate $\varPsi$ into a form $\bigwedge_{p \in P}\left(p \Rightarrow \varPsi_p \right)$ we just need to choose $\varPsi_p = \varPsi\left[p = \TRUE\right]$ obtained from $\varPsi$ by substituting $\TRUE$ to $p$.
Given the inherent property of supporting the $\IDLE$ interaction, the translated conjunctive formula will be equivalent to $\varPsi$ only if the latter allows global idling.
Consider broadcasting from port $p$ to ports $q$ and $r$ (Fig.~\ref{fig:disjunctive-global-constraint}). 
The possible interactions are $p, pq, pr, pqr$ and $\emptyset$ (i.e. idling). 
The disjunctive style formula is: $\neg p \neg q \neg r \vee p \neg q \neg r \vee p q \neg r \vee p \neg q r \vee p q r = \neg q \neg r \vee p$.
The equivalent conjunctive formula is: $(q \Rightarrow p) \wedge (r \Rightarrow p)$ that simply expresses the causal dependency of ports $q$ and $r$ from $p$.

The example below illustrates the application of the two description styles.
\begin{example}[Master-Slaves]
	\label{ex:masterslaves-0}
	Let us consider a simple system consisting of three components: $master$, $slave_1$ and $slave_2$.
	The $master$ performs two sequential requests to $slave_1$ and $slave_2$, and then performs some computation with them.
	
	Figure \ref{fig:masterslaves-0} shows the representation of such components.
	
	\begin{figure}[h]
		\hspace{0.3cm}
		\begin{subfigure}[b]{0.5\textwidth}
			\centering
			\begin{tikzpicture}[->,>=stealth',shorten >=1pt,auto,node distance=2.8cm,
			semithick]
			\tikzstyle{every state}=[fill=white,text=black]
			
			\node[state,double] (A) at (0,1) {$m_{00}$};
			\node[state] (B) at (2,2) {$m_{10}$};
			\node[state] (C) at (2,0) {$m_{01}$};
			\node[state] (D) at (4,1) {$m_{11}$};
			
			\path	(A) edge [bend left]	node[above] {$link_1$}		(B)
			(A) edge [bend right]	node[below] {$link_2$}		(C)
			(B) edge [bend left]	node[above] {$link_2$}		(D)
			(C) edge [bend right]	node[below] {$link_1$}		(D)
			(D) edge	node {$work$}	(A);
			
			\end{tikzpicture}
			\caption{The $master$ component}
		\end{subfigure}
		\begin{subfigure}[b]{0.5\textwidth}
			\centering
			\begin{tikzpicture}[->,>=stealth',shorten >=1pt,auto,node distance=2.8cm,
			semithick]
			\tikzstyle{every state}=[fill=white,text=black]
			
			\node[state,inner sep=5pt,double] (A) at (0,0) {$wait_i$};
			\node[state] (B) at (4,0) {$ready_i$};
			
			\path	(A) edge [bend left]	node {$bind_i$}		(B)
			(B) edge [bend left]	node {$serve_i$}		(A);
			
			\end{tikzpicture}
			\caption{The $slave_i$ component}
		\end{subfigure}
		\caption{$master$ and $slave_i$ components}
		\label{fig:masterslaves-0}
	\end{figure}
	
	The set of allowed interactions $\gamma$ for the set of components $\left\lbrace master, slave_1, slave_2 \right\rbrace$ can be represented via the following \PIL{} formula using the disjunctive style: 
	\begin{align*}
	\varPsi_{disj} = & \left( link_1 \wedge bind_1 \wedge \IDLE[s_2] \right) \vee \left( link_2 \wedge bind_2 \wedge \IDLE[s_1] \right) \vee \\
	& \left( work \wedge serve_1 \wedge serve_2 \right)
	\end{align*}
	where $\IDLE[s_i] \equiv \neg bind_i \wedge \neg serve_i$ is the $\IDLE$ port of $slave_i$.
	
	Alternatively, the same interaction patterns can be modeled using the conjunctive style:
	\begin{align*}
	\varPsi_{conj} = & \left( link_1 \Rightarrow bind_1 \right) \wedge \left( link_2 \Rightarrow bind_2 \right) \wedge \left( bind_1 \Rightarrow link_1 \right) \wedge \left( bind_2 \Rightarrow link_2 \right) \wedge \\ 
	& \left( work \Rightarrow serve_1 \wedge serve_2 \right) \wedge \left(serve_1 \Rightarrow work \right) \wedge \left(serve_2 \Rightarrow work \right)
	\end{align*}
	The two formulas differ in the admissibility of the ``no-interaction'' interaction.
	That is, the conjunctive formula $\varPsi_{conj}$ allows all the components to not interact by performing a transition over their $\IDLE$ ports, while $\varPsi_{disj}$ does not.
	To allow it in the disjunctive case, we could instead consider the following: \begin{align*}
	\varPsi'_{disj} =  \varPsi_{disj} \vee \IDLE[m] \wedge \IDLE[s_1] \wedge \IDLE[s_2]
	\end{align*}
	where $\IDLE[m] \equiv \neg link_1 \wedge \neg link_2 \wedge \neg work$.
	
\end{example}
	
	\section{Static architectures with transfer of values - the \PILOPS{} coordination language}
\label{sec:pilops-framework}

We expand the \PIL{} framework by allowing data exchange between components.
In order to do so, the definition of component will be extended with local variables and the coordination constraints will be expressed with \PILOPS{}, which expands \PIL{} to a notation that is inspired by guarded commands.
Finally we extend the definitions for disjunctive and conjunctive styles and study possible connections between the two.

\subsection{\PILOPS{} components}
\label{subsec:pilops-components}

\begin{definition}[\PILOPS{} Component]\label{def:pilops-component}
	Let $\mathcal{S}$ be the set of all component control locations, $\mathcal{X}$ the set of all local variables, and $\mathcal{P}$ the set of all ports.
	A component is a transition system $B := (S, X, P, T)$, where:
	\begin{itemize}
		\item $S \subseteq \mathcal{S}$: finite set of control locations;
		\item $X \subseteq \mathcal{X}$: finite set of local variables;
		\item $P \subseteq \mathcal{P}$: finite set of ports;
		\item $T \subseteq S \times P \cup \left\lbrace \IDLE \right\rbrace \times S$: finite set of transitions.
		Each transition $(s,p,s')$ can also be denoted by $s \xrightarrow{p} s'$, where $p \in P$ is the port offered for interaction, and such that each transition is labelled by a different port.
	\end{itemize}
	Every component has a special port $\IDLE \notin P$ that is associated to implicit loop transitions $\lbrace s \xrightarrow{idle} s \rbrace_{s \in S}$. 
	
	Furthermore we assume that the sets of ports, local variables and control locations of different components are disjoint.
\end{definition}
A system is a set of coordinated components $B_i = \left( S_i, X_i, P_i, T_i \right)$ for $i=\left[1,n\right]$.
The \emph{configuration} $\Gamma$ of a system is still described by the set of the current control locations of each constituent component, but now it also includes the \emph{valuation function} $\sigma : \mathcal{X} \mapsto \VALUES$ mapping local variables to values:
\begin{align}
\Gamma = \left( \left\lbrace s_i \in S_i \right\rbrace_{i = \left[1..n\right]} , \sigma \right)
\end{align}
Interactions are still sets of ports belonging to different components.

Using a term of the \PILOPS{} language to compose components, the corresponding system configuration $\Gamma$ evolves to a new configuration $\Gamma'$ by performing an interaction $a$ and a set of operations $\Delta$, which we represent with the notation $\Gamma \xrightarrow{a,\Delta} \Gamma'$.

\subsection{Propositional Interaction Logic with Operations (\PILOPS{})}
\label{subsec:pilops-rules}

Let $\mathcal{P}$, $\mathcal{X}$ and $\mathcal{S}$ respectively be the domains of ports, local variables and control locations.
The terms of $\PILOPS(\mathcal{P},\mathcal{X},\mathcal{S})$ are defined by the following syntax:
\begin{align}\label{eq:pilops-syntax}
\text{(\PILOPS{} term)} \quad & \varPhi ::= \RULE{\varPsi}{\Delta} \:|\: \varPhi_1 \ \& \ \varPhi_2 \:|\: \ \varPhi_1 \parallel \varPhi_2 \nonumber \\
\text{(\PIL{} formula)} \quad & \varPsi ::= p \in \mathcal{P} \: | \: \pi \: | \: \NOT{\varPsi} \:|\: \varPsi_1 \wedge \varPsi_2 \nonumber \\
\text{(set of ops.)} \quad & \Delta ::= \emptyset \:|\: \left\lbrace \delta \right\rbrace \:|\: \Delta_1 \cup \Delta_2
\end{align}
where:
\begin{itemize}
	\item operators $\&$ and $\parallel$ are \emph{associative} and \emph{commutative}, with $\&$ having higher precedence than $\parallel$;
	\item $\pi: 2^\Gamma \mapsto \left\lbrace \TRUE,\FALSE \right\rbrace$ is a state predicate;
	\item $\delta: 2^\sigma \mapsto 2^\sigma$ is an operation that transforms the valuation function $\sigma$.
\end{itemize}

The models of the logic are still interactions $a$ on $\mathcal{P}$, where the satisfaction relation is defined by the set of rules (\ref{eq:pil-sat-relation}) for \PIL{} with the following extension:
\begin{alignat}{2}\label{eq:pilops-sat-relation}
\MODELS{a}{\Gamma}{&\RULE{\varPsi}{\Delta}} &\quad&\text{if } \MODELS{a}{\Gamma}{\varPsi} \nonumber \\
\MODELS{a}{\Gamma}{&\varPhi_1 \ \& \ \varPhi_2} &&\text{if } \MODELS{a}{\Gamma}{\varPhi_1} \text{ and } \MODELS{a}{\Gamma}{\varPhi_2} \nonumber \\
\MODELS{a}{\Gamma}{&\varPhi_1 \parallel \varPhi_2} &&\text{if } \MODELS{a}{\Gamma}{\varPhi_1} \text{ or } \MODELS{a}{\Gamma}{\varPhi_2}  
\end{alignat}
In other words, the conjunction and disjunction operators $\&$ and $\parallel$ for \PILOPS{} terms are equivalent to the logical $\wedge$ and $\vee$ from the interaction semantics perspective.

Operations in $\Delta$ are treated in a different way: operations associated to rules combined with ``$\&$'' will be either performed all together if the associated \PIL{} formulas hold for $a,\Gamma$ or not at all if at least one formula does not, while for rules combined with the ``$\parallel$'' operator a maximal union of operations satisfying the \PIL{} formulas will be executed.

We indicate the set of operations to be performed for $\varPhi$ under $a,\Gamma$ as $\EXECOPS{\varPhi}$, which is defined according to the following rules:
\begin{alignat}{2}\label{eq:pilops-ops-function}
\EXECOPS{\RULE{\varPsi}{\Delta}} =&\begin{cases}
\Delta & \text{if } \MODELS{a}{\Gamma}{\varPsi}\\
\emptyset & \text{otherwise}
\end{cases} \nonumber \\
\EXECOPS{\varPhi_1 \ \& \ \varPhi_2} =&\begin{cases}
\EXECOPS{\varPhi_1} \cup \EXECOPS{\varPhi_2} & \text{if } \MODELS{a}{\Gamma}{\varPhi_1} \text{ and } \MODELS{a}{\Gamma}{\varPhi_2} \\
\emptyset & \text{otherwise}
\end{cases} \nonumber \\
\EXECOPS{\varPhi_1 \parallel \varPhi_2} =& \ \EXECOPS{\varPhi_1} \cup \EXECOPS{\varPhi_2} 
\end{alignat}
Two \PILOPS{} terms $\varPhi_1,\varPhi_2$ are \emph{equivalent} if, for any interaction $a$ and configuration $\Gamma$, $\EXECOPS{\varPhi_1} = \EXECOPS{\varPhi_2}$.

\subsubsection{Axioms for \PILOPS{}}

The following axioms hold for \PILOPS{} terms:
\begin{align}
&\& \ \text{is associative, commutative and idempotent} \\
&\RULE{\varPsi_1}{\Delta_1} \ \& \ \RULE{\varPsi_2}{\Delta_2} = \RULE{\varPsi_1 \wedge \varPsi_2}{\Delta_1 \cup \Delta_2} \\
&\varPhi \ \& \ \RULE{\TRUE}{\emptyset} = \varPhi \\
&\parallel \ \text{is associative, commutative and idempotent}\\
&\RULE{\varPsi_1}{\Delta} \parallel \RULE{\varPsi_2}{\Delta} = \RULE{\varPsi_1 \vee \varPsi_2}{\Delta}\\
&\RULE{\varPsi}{\Delta_1} \parallel \RULE{\varPsi}{\Delta_2} = \RULE{\varPsi}{\Delta_1 \cup \Delta_2}\\
&\RULE{\FALSE}{\Delta} \parallel \varPhi = \varPhi\\
&\text{Absorption:} \ \varPhi_1 \parallel \varPhi_2 = \varPhi_1 \parallel \varPhi_2 \parallel \varPhi_1 \ \& \ \varPhi_2 \label{eq:absorption-axiom}\\
&\text{Distributivity:} \ \varPhi \ \& \left( \varPhi_1 \parallel \varPhi_2 \right) = \varPhi \ \& \ \varPhi_1 \parallel \varPhi \ \& \ \varPhi_2 \label{eq:distributivity-axiom}\\
&\text{Normal disjunctive form (DNF):}  \\ 
& \RULE{\varPsi_1}{\Delta_1} \parallel \RULE{\varPsi_2}{\Delta_2} = \RULE{\varPsi_1 \wedge \NOT{\varPsi_2}}{\Delta_1} \parallel \RULE{\varPsi_2 \wedge \NOT{\varPsi_1}}{\Delta_2} \parallel \RULE{\varPsi_1 \wedge \varPsi_2}{\Delta_1 \cup \Delta_2} \nonumber
\end{align}
Note that \PILOPS{} strictly contains \PIL{} as a formula $\varPsi$ can be represented by $\RULE{\varPhi}{\emptyset}$.
The operation $\&$ is the extension of conjunction with neutral element $\RULE{\TRUE}{\emptyset}$ and $\parallel$ is the extension of the disjunction with an absorption (\ref{eq:absorption-axiom}) and distributivity axiom (\ref{eq:distributivity-axiom}).
The DNF is obtained by application of the axioms.
Note some important differences with \PIL{}: the usual absorption axioms for disjunction and conjunction are replaced by a single absorption axiom (\ref{eq:absorption-axiom}) and there is no conjunctive normal form.

\subsubsection{Operations}

Operations $\delta$ in \PILOPS{} are assignments on local variables of components involved in an interaction of the form $\ASSIGN{}{x}{f}$, where $x \in \mathcal{X}$ is the local variable subject to the assignment and $f : \mathsf{V}^k \mapsto \mathsf{V}$, is a function on local variables $y_1,\dots,y_k$ ($y_i \in \mathcal{X}$) on which the assigned value depends.

We can define the semantics of the application of the assignment $x := f$ to the valuation function $\sigma$ as:
\begin{align}
\left(\ASSIGN{}{x}{f}\right) \left(\sigma\right) = \sigma\left[x \mapsto f\left(\sigma \left(y_1\right),\dots,\sigma \left(y_k\right)\right)\right]
\end{align}

A set of assignment operations $\Delta$ is performed using a \emph{snapshot semantics}. 
When $\Delta$ contains multiple assignments on the same local variable, the results are \emph{non-deterministic}.

A \PILOPS{} term $\varPhi$ is a coordination mechanism that, applied to a set of components $B_1 \dots B_n$, gives a system defined by the following rule:

\begin{align}\label{eq:pilops-evolution-semantics}
\inference{
	\MODELS{a}{\Gamma}{\varPhi} \qquad
	\forall p \in a: s_i \xrightarrow{p} s'_i \qquad
	\sigma' \in \EXECOPS{\varPhi} \left( \sigma \right)
}{
	\left( \left\lbrace s_i \right\rbrace_{\left[1..n\right]}, \sigma \right) \xrightarrow{a} \left(\left\lbrace s'_i \right\rbrace_{\left[1..n\right]}, \sigma'\right)
}[]
\end{align}
where $\EXECOPS{\varPhi} \left( \sigma \right)$ is the set of valuation functions obtained by applying the operations $\delta \in \EXECOPS{\varPhi}$ to the valuation function $\sigma$ in every possible order (using a \emph{snapshot semantics}).

\subsection{Disjunctive vs. conjunctive specification style in \PILOPS{}}
\label{subsec:pilops-styles}

We define disjunctive and conjunctive style specification in \PILOPS{}. 
We associate with $p \Rightarrow \varPsi_p$ an operation $\Delta_p$ to be performed when an interaction involving $p$ is executed according to this rule. 
We call the \PILOPS{} term  describing this behavior the \emph{conjunctive term} $\irRULE{}{p}{\varPsi_p}{\Delta_p} = \left( \RULE{\neg p}{\emptyset} \parallel \RULE{p \wedge \varPsi_p}{\Delta_p} \right)$.
$\Delta_p$ may be executed when $p$ is involved in some interaction; otherwise, no operation is executed. 

The conjunction of terms of this form gives a disjunctive style formula. 
Consider for instance, the conjunction of two terms: 
{\small\begin{align*}
	&\irRULE{}{p}{\varPsi_p}{\Delta_p} \& \irRULE{}{q}{\varPsi_q}{\Delta_q} = \left( \RULE{\neg p}{\emptyset} \parallel \RULE{p \wedge  \varPsi_p}{\Delta_p}\right) \& \left( \RULE{\neg q}{\emptyset} \parallel \RULE{q \wedge  \varPsi_q}{\Delta_q} \right) = \\
	&=\RULE{\neg p \wedge \neg q}{\emptyset} \parallel \RULE{p \wedge \neg q \wedge  \varPsi_p}{\Delta_p} \parallel \RULE{q \wedge \neg p \wedge \varPsi_q}{\Delta_q} \parallel \RULE{p \wedge q \wedge \varPsi_p \wedge \varPsi_q}{\Delta_p \cup \Delta_q}
	\end{align*}}
The disjunctive form obtained by application of the distributivity axiom (\ref{eq:distributivity-axiom}) is a union of four terms corresponding to the canonical monomials on $p$ and $q$ and leading to the execution of no operation, either operation $\Delta_p$, $\Delta_q$ or both.
It is easy to see that for a set of ports $P$ the conjunctive form 
\begin{equation*}
\bigand_{p \in P} \left(\RULE{\neg p}{\emptyset} \parallel \RULE{p \wedge \varPsi_p}{\Delta_p} \right)
\end{equation*}
is equivalent to the disjunctive form 
\begin{equation*}
\bigparallel_{I \cup J = P} \Big( \bigwedge_{i \in I} \RULE{p_i \wedge \varPsi_{p_i} \bigwedge_{j \in J} \neg p_j}{\bigcup_{i \in I} \Delta_{p_i}} \Big)
\end{equation*}
where $\bigcup_{p_i \in \emptyset} \Delta_{p_i} = \emptyset$.

The converse does not hold. 
Given a disjunctive specification it is not always possible to get an equivalent conjunctive one. 
If we have a term of the form $\bigparallel_{k \in K} \RULE{\varPsi}{\Delta_k}$ over a set of ports $P$, it can be put in canonical form and will be the union of canonical terms of the form $\bigwedge_{i \in I} \RULE{p_i \bigwedge_{j \in J} \neg p_j}{\Delta_{IJ}} $.
It is easy to see that for this form to be obtained as a conjunction of causal terms a sufficient condition is that for each port $p_i$ there exists an operation $\Delta_{p_i}$ such that $\Delta_{IJ} = \bigcup_{i \in I} \Delta_{p_i}$.
That is, the operation associated with a port participating to an interaction is the same. 
This condition also determines the limits of the conjunctive and compositional approach. 

\begin{example}[Master-Slaves]
	\label{ex:masterslaves-1}
	Let us expand the example scenario introduced in Example \ref{ex:masterslaves-0} by attaching data transfer between the $master$ component and the two $slave_1$ and $slave_2$ components.
	More specifically, we assume that the $master$ has a $buffer$ local variable that will take the value obtained by adding the values stored in local variables $mem_1$ and $mem_2$ of the two respective slaves when they all synchronize through the ports $work,serve_1,serve_2$.
	
	The set of allowed interactions $\gamma$ is not going to change, but adopting the \PILOPS{} coordination language we can characterize the desired behaviour using the disjunctive style as follows: 
	{\small\begin{align*}
		\varPhi_{disj} = \ & \RULE{link_1 \wedge bind_1 \wedge \IDLE[2]}{\emptyset} \parallel
		\RULE{link_2 \wedge bind_2 \wedge \IDLE[1]}{\emptyset} \parallel \\
		& \RULE{work \wedge serve_1 \wedge serve_2}{buffer := mem_1 + mem_2}
		\end{align*}}
	The conjunctive style version equivalent to $\varPhi_{disj}$ (except for its allowance of the idling of all components) is the following:
	{\small\begin{align*}
		\varPhi_{conj} = &\irRULE{}{link_1}{bind_1}{\emptyset} \& \irRULE{}{link_2}{bind_2}{\emptyset} \&
		\irRULE{}{bind_1}{link_1}{\emptyset} \& \irRULE{}{bind_2}{link_2}{\emptyset} \& \\ 
		& \irRULE{}{work}{serve_1 \wedge serve_2}{\ASSIGN{}{buffer}{mem_1+mem_2}} \&\\
		&  \irRULE{}{serve_1}{work}{\emptyset} \& \irRULE{}{serve_2}{work}{\emptyset}
		\end{align*}}
	
\end{example}
	
	\section{The \LANG{} framework}
\label{sec:full-framework}

%In Section~\ref{sec:pilops-framework} we presented an extension to the minimal framework described in Section~\ref{sec:pil-framework} that encompasses data exchange.

%In this Section we will present the full \LANG{} framework, an extension to \PILOPS{} allowing dynamic creation/deletion of components, structuring the system architecture into \emph{motifs} that provide specific coordination laws to attached components, and capturing mobility properties by managing (dynamic) data structures that act as ``maps'' for components.
In this Section we present the DReAM framework, allowing dynamism and
reconfiguration which extends the static framework in the following manner.
Components are instances of types of components and their number can dynamically change. Coordination between components in a motif, but also between the motifs constituting a system, is expressed by the DReAM coordination language, a first order extension of PILOps. In motifs coordination is parametrized by the notion of map which is an abstract relation used as a reference to model topology of the underlying architecture as well as component mobility.
%In this Section we present the full \LANG{} framework, allowing dynamism and reconfiguration. 
%A system is built by composing instances of component types. 
%It is structured in motifs that are architectures characterized by specific coordination and reconfiguration rules. 
%Interactions between motifs may involve migration of components.
%
%The \LANG{} framework extends the static framework in the following manner. 
%Components are instances of types of components and their number can dynamically change. 
%Coordination between components in a motif but also between the motifs constituting a system, is expressed by the \LANG{} coordination language, a first order extension of \PILOPS{}. 
%In motifs coordination is parametrized by maps which is an abstract relation used as a reference to model topology of the underlying architecture as well as component mobility.
%
%We present basic definitions about components, the \LANG{} coordination language, motifs and systems with their underlying operational semantics.

\subsection{Component Types and component Instances}
\label{subsec:components}

\LANG{} systems are constituted by \emph{instances of component types}.
Component types in \LANG{} correspond to \PILOPS{} components (see Definition~\ref{def:pilops-component}), while component instances are obtained from a component type by renaming its control locations, ports and local variables with a unique \emph{identifier}.

%\begin{definition}[Component type]
%	Let $\mathcal{S}$ be the set of all component control locations and $\mathcal{P}$ the set of all ports.
%	A component type is a transition system $b := (S, X, P, T)$, where:
%	\begin{itemize}
%		\item $S \subseteq \mathcal{S}$: finite set of control locations;
%		\item $X \subseteq \mathcal{X}$: finite set of variables;
%		\item $P \subseteq \mathcal{P}$: finite set of ports;
%		\item $T \subseteq S \times P \cup \left\lbrace \IDLE \right\rbrace \times S$: finite set of transitions.
%		Each transition $(s,p,s')$ can also be denoted by $s \xrightarrow{p} s'$, where $p \in P$ is the port offered for interaction, and such that each transition is characterized by a different port.
%	\end{itemize}
%	Notice also that every component type has a special port $idle \in P$ with transitions $\lbrace s \xrightarrow{idle} s \rbrace_{s \in S}$. 
%	Furthermore we assume that the sets of ports, states and variables of different component types are disjoint.
%\end{definition}

To highlight the relationships between component types and their defining sets we use a ``dot notation'':
\begin{itemize}
	\item $b.S$ refers to the set of control locations $S$ of component type $b$ (same for ports and variables);
	\item $b.s$ refers to the control location $s\in b.S$ (same for ports and variables).
\end{itemize}

\begin{definition}[Component instance]
	Let $\mathcal{C}$ be the domain of instance identifiers $\mathcal{C}$ and $B = \left\langle b_1, \dots , b_n \right\rangle$ be a tuple of component types where each element is $b_i = (S_i, X_i, P_i, T_i)$.
	
	A set of \emph{component instances} of type $b_i$ is represented by $b_i.C = \left\lbrace b_i.c : c \in C \right\rbrace$, for $1 \leq i \leq n$ and $C \subseteq \mathcal{C}$, and is obtained by renaming the set of control locations, ports and local variables of the component type $b_i$ with $c$, that is $b_i.c = (c.S_{i}, c.X_{i}, c.P_{i}, c.T_{i})$.
	Without loss of genericity, we assume that instance identifiers uniquely represent a component instance regardless of its type.
	
	The state of a component instance $b.c$ is therefore defined as the pair $\left\langle c.s,c.\sigma \right\rangle$, where $c.\sigma$ is the \emph{valuation function} of the variables $c.X$\footnote{Notice that when writing e.g. $c.s$ we are omitting the explicit reference to the component type $b$ and using a shorter notation compared to the complete one, e.g. $b.c.s$.}.
	We use the same notation to denote ports, states and variables belonging to a given component instance (e.g. $c.p \in c.P$) and assume that ports of different component instances are still disjoint sets, i.e. $c.P \cap c'.P = \emptyset$ for $c \neq c'$.
\end{definition}

%We assume that ports of different component instances are still disjoint sets, i.e. $c.P \cap c'.P = \emptyset$ for $c \neq c'$.

Transitions for component instances $c.T$ are obtained from the respective component type transitions $T$ via port name substitution, i.e. via the rule:
\begin{equation}
\label{eq:transition-rename}
\inference{(s,p,s') \in T}{c.s \xrightarrow{c.p} c.s'}[]
\end{equation}

\subsection{The \LANG{} coordination language}
\label{subsec:evolution-rules}

The \LANG{} coordination language is essentially a first-order extension of \PILOPS{} where quantification over sets of components is introduced.

Given the domain of ports $\mathcal{P}$, the \LANG{} coordination language is defined by the syntax:
\begin{align}\label{eq:foilops-syntax}
\text{(\LANG{} term)} \quad & \rho ::= \varPhi \:|\: \evRULE{D}{\varPhi} \:|\: \rho_1 \ \& \ \rho_2 \:|\: \rho_1 \parallel \rho_2 \nonumber \\
\text{(declaration)} \quad & D ::= \forall c : m.b \:|\: \exists c : m.b \:|\: D_1 , D_2 \nonumber \\
\text{(\PILOPS{} term)} \quad & \varPhi ::= \RULE{\varPsi}{\Delta} \:|\: \varPhi_1 \ \& \ \varPhi_2 \:|\: \ \varPhi_1 \parallel \varPhi_2 \nonumber \\
\text{(\PIL{} formula)} \quad & \varPsi ::= c.p \in \mathcal{P} \: | \: \pi \: | \: \NOT{\varPsi} \:|\: \varPsi_1 \wedge \varPsi_2 \nonumber \\
\text{(set of ops.)} \quad & \Delta ::= \emptyset \:|\: \left\lbrace \delta \right\rbrace \:|\: \Delta_1 \cup \Delta_2
\end{align}
%where:
\begin{itemize}
	\item \emph{Declarations} define the context of the term by declaring quantified ($\forall | \exists$) component variables ($c$) associated to instances of a given type ($b$) belonging to a motif $m$;
	\item Operators $\&$ and $\parallel$ are the same as the ones introduced in (\ref{eq:pilops-syntax}) for \PILOPS{};
	\item $\pi: 2^\Gamma \mapsto \left\lbrace \TRUE,\FALSE \right\rbrace$ is a state predicate on the system configuration $\Gamma$;
	\item $\delta: 2^\Gamma \mapsto 2^\Gamma$ is an \emph{operation} that transforms the system configuration $\Gamma$.
\end{itemize}

A \LANG{} coordination term is \emph{well formed} if its \PIL{} formulas and associated operations contain only component variables that are defined in its declarations.
From now on, we will only consider well formed terms.

Given a system configuration, a coordination term can be translated to an equivalent \PILOPS{} term by performing a \emph{declaration expansion} step, by expanding the quantifiers and replacing component variables with actual components.
%Indeed, a \LANG{} term without declarations is equivalent to a \PILOPS{} term.

%We will now detail each step that allows to perform this transformation in order to formally describe the evolution semantics of motifs.

\subsubsection{Declaration expansion for coordination terms}
\label{subsubsec:ev-rules-concretization}

Given that \LANG{} systems host finite numbers of component instances, first-order logic quantifiers can be eliminated by enumerating every component instance of the type specified in the declaration.
We thus define the \emph{declaration expansion} $\CONCRETE{\rho}$ of $\rho$ under configuration $\Gamma$ via the following rules:
\begin{equation}\label{eq:ev-concretization}
\begin{aligned}[c]
\CONCRETE{\varPhi} = &\ \varPhi & \CONCRETE{\evRULE{\forall c:m.b}{\varPhi}} = & \bigand_{c^{*} \in m.b.C} \varPhi \left[c^{*} / c\right] \\
\CONCRETE{\rho_1 \ \& \ \rho_2} = & \CONCRETE{\rho_1} \ \& \ \CONCRETE{\rho_2} & \CONCRETE{\evRULE{\exists c:m.b}{\varPhi}} = & \bigparallel_{c^{*} \in m.b.C} \varPhi \left[c^{*} / c\right]\\
\CONCRETE{\rho_1 \parallel \rho_2} = & \CONCRETE{\rho_1} \parallel \CONCRETE{\rho_2} & \CONCRETE{\evRULE{D_1 , D_2}{\varPhi}} = & \CONCRETE{\evRULE{D_1}{\CONCRETE{\evRULE{D_2}{\varPhi}}}}
\end{aligned}
\end{equation}
%\begin{align}\label{eq:ev-concretization}
%\CONCRETE{\varPhi} = &\ \varPhi \nonumber \\
%\CONCRETE{\rho_1 \ \& \ \rho_2} = & \CONCRETE{\rho_1} \ \& \ \CONCRETE{\rho_2} \nonumber \\
%\CONCRETE{\rho_1 \parallel \rho_2} = & \CONCRETE{\rho_1} \parallel \CONCRETE{\rho_2} \nonumber \\
%\CONCRETE{\evRULE{\forall c:b}{\varPhi}} = & \bigand_{c^{*} \in b.C} \varPhi \left[c^{*} / c\right] \nonumber\\
%\CONCRETE{\evRULE{\exists c:b}{\varPhi}} = & \bigparallel_{c^{*} \in b.C} \varPhi \left[c^{*} / c\right] \nonumber\\
%\CONCRETE{\evRULE{D_1 , D_2}{\varPhi}} = & \CONCRETE{\evRULE{D_1}{\CONCRETE{\evRULE{D_2}{\varPhi}}}}
%\end{align}
where $m.b.C$ is the set of component instances of type $b$ in motif $m$, and $\left[c^{*} / c\right]$ is the substitution of the symbol $c$ with the actual identifier $c^{*}$ in the associated term.

By applying (\ref{eq:ev-concretization}), any  term can be transformed into a \PILOPS{} term, whose semantics is defined in Section~\ref{subsec:pilops-rules}:
%This means that, at any given configuration, the semantics of a coordination term are equivalent to the semantics of its \PILOPS{}-equivalent obtained through declaration expansion.
%Therefore, the models of the logic for coordination terms are unchanged from what has been presented in Section~\ref{subsec:pilops-rules}:
%\begin{itemize}
%	\item the satisfaction relation is the one defined by (\ref{eq:pil-sat-relation}) and (\ref{eq:pilops-sat-relation});
%	\item the set of operations to be performed is defined according to (\ref{eq:pilops-ops-function}).
%\end{itemize}

\subsection{Motif modeling}
\label{subsec:motif}

A motif characterizes an independent dynamic architecture involving a set of component instances $C$ subject to specific \emph{coordination terms} parameterized by a specific data structure called \emph{map}.
\begin{definition}[Motif]\label{def:motif}
	Let $\mathcal{C}$ be the domain of component instance identifiers.
	A \emph{motif} is a tuple $m := \left\langle C, \rho, Map_0, \MVAt_0 \right\rangle$, where $C \subseteq \mathcal{C}$ is the set of component instances assigned to the motif, $\rho$ is the coordination term regulating interactions and reconfigurations among them, and $Map_0,\MVAt_0$ are the initial configurations of the map associated to the motif and of the addressing function.
	
	We assume that each component instance is associated with exactly one motif, i.e. $m_1.C \cap m_2.C = \emptyset$.
\end{definition}
A $Map$ is a set of locations and a connectivity relation between them.
It is the structure over which computation is distributed and defines a system of coordinates for components.
It can represent a physical structure e.g. geographic map or some conceptual structure,
% used to decompose the problem of system design 
e.g., cellular structure of a memory.
In \LANG{} a map is specified as a graph $Map = \left(N,E\right)$, where:
\begin{itemize}
	\item $ N $ is a set of nodes or locations (possibly infinite);
	\item $ E $ is a set of edges subset of $N \times N$ that defines the connectivity relation between nodes.
\end{itemize}

The relation $E$ defines a concept of neighborhood for components. 
%If the memory is empty then the only available information for a location is its name. Otherwise, 
%The memory can be shared by different components and used for their coordination.
%The relation $ E $ defines a concept of neighborhood that in many applications is used to express coordination constraints or directions for moving components: if $ E $ is empty, then it means that the map is purely an indexing structure. 
%On the other hand, if the memory is empty then it means that the only available information for a location is its name. 
%Otherwise, the memory being shared by different components can be used for their coordination, like in the thread-based model.\\

Component instances $C$ in a motif and its map are related through the (partial) \emph{address function} $\MVAt : \mathcal{C} \rightarrow N$ binding each component in $C$ to a node $n \in N$ of the map. 
%(e.g., $\MVAt(c)=n$ means that component instance $c$ is associated to location $n$ in the map).
%
%Conversely we can define the inverse of $\MVAt$ as the \emph{content function} $\sharp : N \rightarrow 2^C$ binding each node $n \in N$ of the map to the set of instances $C' \subseteq C$ which are located in $n$.
%Notice that the adopted definitions for the address and content functions implies that more components can be located in the same node, but at no point a component can be bound to two different nodes.\\
%
%For instance if a map is an array then $ N=\left\lbrace 1,2, \dots, i, \dots \right\rbrace $ and $ E=(i,i+1) $ for $ i \in N $. 
%A pipeline is a system such that for any components $ b $ and $ b' $ such that $ \MVAt(b')=\MVAt(b)+1 $  the input of $ b' $ is connected to the output of $ b $.

Maps can be used to model a physical environment where components are moving.
% evolving - for instance - in the form of 
If the map is an array $ N= \left\lbrace (i,j) | i,j \in \mathsf{Integers}\right\rbrace \times \left\lbrace f,o \right\rbrace $, the pairs 
$(i,j)$ represent coordinates and the symbols $ f $ and $ o $ stand respectively for free and obstacle.
We can model the movement of $b$
%express the fact that a component $ b $ 
such that $ \MVAt(b) = ((i,j), f) $ 
to a position $ (i+a, j+b) $ provided that there is a path from $ (i,j) $ to $ (i+a, j+b) $ consisting of free cells.

The \emph{configuration} $\Gamma_m$ of motif $m$ is represented by the tuple
\begin{align}
\Gamma_m =&  \MOTIFconf{C}{}{}{}{}\\
\equiv &   \left\langle \left\lbrace c.s \right\rbrace_{c \in m.C}, \left\lbrace c.\sigma \right\rbrace_{c \in m.C}, m.Map , m.\MVAt \right\rangle
\end{align}
%The configuration of a motif changes according to its \emph{coordination rules} by performing transitions labelled with an \emph{interaction} and a \emph{sequence of operations}.
%We distinguish coordination rules of motifs in \emph{interaction rules} - i.e. rules that describe how components interact and exchange data - and \emph{configuration rules} - i.e. rules that characterize the dynamic aspects of the motif such as components creation/deletion, changes in the map and in the address function.

%We will also refer to the global valuation function of all component instances with $\Sigma$:
%\begin{equation}
%\Sigma = \bigcup_{c\in C} c.\sigma
%\end{equation}

%Our definition of motif provides a basis for the taxonomy of different types of dynamism.
%In the function $ \MVAt : C \rightarrow N $ which characterizes a motif we can change:
By modifying the configuration of a motif we can model:
\begin{itemize}
	\item \emph{Component dynamism}: The set of component instances $ C $ may change by creating/deleting or migrating components;
	\item \emph{Map dynamism}: The set of nodes or/and the connectivity relation of a map may change.
	This is the case in particular when an autonomous component e.g. a robot, explores an unknown environment and builds a model of it;
	\item \emph{Mobility dynamism}: The address function $ \MVAt $ changes to express mobility of components.
	%when component coordinates change because they are moving in a geographic map or in a logically-defined structure, e.g. threads in a memory space.
\end{itemize}
Different types of dynamism can be obtained as the combination of these three basic types. 
%Component dynamism is very common in computing systems such as multiprocessor systems where the number of processes may change depending on the needs. 
%Map dynamism is needed to model systems with dynamically changing environment but also for computing systems where the underlying computation structure changes. 
%Finally, mobility is a very common feature of autonomous  systems e.g. autonomous vehicles and swarm robotics.

\subsubsection{Reconfiguration operations}

Reconfiguration operations realize component, map and mobility dynamism by allowing transformations of a motif configuration at runtime.\\
Component dynamism can be realized using the following statements:
\vspace{-0.5em}
\begin{itemize}
	\item $\CREATE{b}{n}$: creates an instance of type $b$ at node $n$ of the relevant map;
	\item $\DELETE{c}$: deletes instance $c$.
\end{itemize}
Map dynamism can be realized using the following statements:
\vspace{-0.5em}
\begin{itemize}
	\item $\ADDNODE{n}$: adds node $n$ to the relevant map;
	\item $\REMOVENODE{n}$: removes node $n$ from the relevant map, along with incident edges and components mapped to it;
	\item $\ADDEDGE{n_1}{n_2}$: adds edge $\left(n_1,n_2\right)$ to the relevant map;
	\item $\REMOVEEDGE{n_1}{n_2}$: removes edge $\left(n_1,n_2\right)$ from the relevant map.
\end{itemize}
Mobility dynamism can be realized using the following statement:
\vspace{-0.5em}
\begin{itemize}
	\item $\MOVE{c}{n}$: changes the position of $c$ to node $n$ in the relevant map.
	%	\item $\MIGRATE{c}{m}{n}$: moves $c$ to node $n$ in the map of motif $m$.
\end{itemize}

\subsubsection{Operational semantics of motifs}Terms $\rho$ of the coordination language are used to compose component instances in a motif. 
The latter can evolve from a configuration $\Gamma_m$ to another $\Gamma''_m$ by performing a transition labelled with the interaction $a$ and characterized by the application of the set of operations $\EXECOPS[a,\Gamma_m]{\CONCRETE[\Gamma_m]{\rho}}$ iff $a \models \CONCRETE[\Gamma_m]{\rho}$.
Formally this is encoded by the following inference rule:
\begin{align}\label{eq:ev-semantics-1}
\inference{
	\MODELS{a}{\Gamma_m}{\CONCRETE[\Gamma_m]{\rho}} \qquad 
	%	\left\lbrace c.s \right\rbrace_{c \in C} \xrightarrow{a} \left\lbrace c.s' \right\rbrace_{c \in C} \qquad
	\Gamma_m \xrightarrow{a} \Gamma'_m \qquad
	%	\MOTIFconf[]{C'}{'}{'}{'}{'} = \EXECOPS{\CONCRETE{\rho}} \left( \MOTIFconf[]{C}{}{}{}{} \right)
	\Gamma''_m \in \EXECOPS[a,\Gamma_m]{\CONCRETE[\Gamma_m]{\rho}} \left( \Gamma'_m \right)
}{
	%	\MOTIFconf[]{C}{}{}{}{} \xrsquigarrow[5]{a} \MOTIFconf[]{C'}{'}{'}{'}{'}
	\Gamma_m \xrsquigarrow[5]{ \ \ a \ \ } \Gamma''_m
}[]
\end{align}
%where:
\begin{itemize}
	\item $\Gamma_m \xrightarrow{a} \Gamma'_m$ expresses the capability of the motif to evolve to a new configuration through interaction $a$ according to the simple \PIL{} semantics of (\ref{eq:pil-op-semantics-2}).
	By expanding the motif configuration we have indeed:
	\begin{align}\label{eq:interaction-ev-semantics}
	\inference{
		\forall c.p \in a: c.s \xrightarrow{c.p} c.s' \quad \text{with } c \in m.C
	}{
		\MOTIFconf{C}{}{}{}{} \xrightarrow{a} \MOTIFconf{C}{'}{}{}{}
	}[]
	\end{align}
	\item $\EXECOPS[a,\Gamma_m]{\CONCRETE[\Gamma_m]{\rho}} \left( \Gamma'_m \right)$ is the set of motif configurations obtained by applying the operations $\delta \in \EXECOPS[a,\Gamma_m]{\CONCRETE[\Gamma_m]{\rho}}$ in every possible order (evaluated using a snapshot semantics).
\end{itemize}

\subsection{System-level operational semantics}

\begin{definition}[\LANG{} system]\label{def:foilops-system}
	Let $B$ be a tuple of component types and $M$ a set of motifs.
	A \emph{\LANG{} system} is a tuple $\left\langle B,M,\mu,\Gamma_0 \right\rangle$ where $\mu$ is a \emph{migration term} and $\Gamma_0$ is the initial configuration of the system.
\end{definition}
The migration term $\mu$ is a coordination term where the operations $\delta$ are of the form $\MIGRATE{c}{m}{n}$, which move a component instance $c$ to node $n$ in the map of motif $m$.

The global configuration of a \LANG{} system is simply the union of the configurations of the set of motifs $M$ that constitute it:
\begin{align}
\Gamma  = &  \bigsqcup_{m \in M} \Gamma_m = \left\langle \bigcup_m m.C.s , \bigcup_m m.C.\sigma , \bigcup_m m.Map, \bigcup_m m.\MVAt \right\rangle
\end{align}
where we overloaded the semantics of the union operator to combine different maps in a bigger one characterized by the union of the sets of nodes, edges and memory locations.

The system-level semantics is described by the following inference rule:
\begin{align}\label{eq:system-semantics}
\inference{
	\Gamma_m \xrsquigarrow[5]{ \ \ a_m \ \ } \Gamma'_m \ \text{ for } m \in M \qquad
	\MODELS{a}{\Gamma'}{\CONCRETE[\Gamma']{\mu}} \quad
	\Gamma'' \in \EXECOPS[a,\Gamma']{\CONCRETE[\Gamma']{\mu}} \left( \Gamma' \right)
}{
	\Gamma \xrightarrow{a} \Gamma''
}[]
\end{align}
%where:
\begin{itemize}
	\item $\Gamma' = \bigsqcup_{m \in M} \Gamma'_m$;
	\item $a_m \subseteq a$ is a subset of the global interaction $a$ containing only ports of component instances belonging to motif $m$. 
\end{itemize}
By performing interaction $a$ each motif first evolves on its own according to its coordination term, and then the whole system changes configuration according to the migration term $\mu$.

\subsection{Implementation principle}
\label{subsubsec:implementation}

The ongoing implementation of the \LANG{} framework involves two parts: a Java execution engine with an associated API and a domain-specific language (DSL) with an IDE for modeling in \LANG{}.

The execution engine directly implements the \LANG{} operational semantics.
Components and maps are defined as abstract classes that the programmer can extend with custom functions for which a library of predefined implementations is provided.
Furthermore, by using directly the API, the programmer can enrich coordination terms and associated operations with any Java code.

The DSL implements the abstract syntax of \LANG{} using XText, which also provides an integrated development environment as an Eclipse plugin with convenient features like syntax highlighting and static checks.

Given the dynamic nature of the modeled systems and the importance of the study of collective behaviors, we are also realizing a pluggable component for the execution engine to visualize the evolution of \LANG{} system configurations.

We provide an abstract syntax of \LANG{}, for a system with a set of motifs $M$ (with their respective component instances) and migration term describing how components can leave a motif and join another.

\noindent\textbf{System} \{

$B = \left\lbrace b_1, \dots, b_k \right\rbrace$ (the set of component types)

$M = \left\lbrace m_1, \dots, m_n \right\rbrace$ (the set of motifs)

$\mu$ (migration term)

\noindent\}

\noindent\textbf{Motif} $m_i$ \{

$Map_i$ (definition and associated functions/predicates)

$\rho_i$ (coordination term)

\noindent\}

Both migration and motif terms are expressions built using operators $\&$, $\parallel$ and the following ``basic'' terms:
\begin{align}
\text{conjuctive term:} \;&\evRULE{\forall c:b \in m , D}{\irRULE{}{c.p}{\phi_p}{\Delta_{p}}}  \label{eq:conj-rule}\\
\text{disjunctive term:} \;&\evRULE{D}{\RULE{\phi}{\Delta_{\phi}}} \label{eq:disj-rule}\\
\text{restriction term:} \;&\AtMost{n}{b.p} \label{eq:restriction-rule}
\end{align}
where:
\begin{itemize}
	\item $b \in B$ is a component type in the system;
	\item $D$ is a declaration as defined in (\ref{eq:foilops-syntax});
	\item $\phi_p, \phi$ are \PIL{} formulas;
	\item $\Delta_{p}, \Delta_{\phi}$ are sets of operations;
	\item $n \in \mathbb{N}$ is a integer;
	\item $p \in b.P$ is a port of component type $b$.
\end{itemize}
The conjunctive term (\ref{eq:conj-rule}) matches the one defined for \PILOPS{} in Section \ref{subsec:pilops-styles}.
Its meaning is that any component instance $c$ of type $b$ belonging to motif $m$ interacts through port $p$ if $\phi_p$ holds, and the corresponding operation is $\Delta_{p}$.

The disjunctive term (\ref{eq:disj-rule}) is, in fact, a general \LANG{} coordination term.
It characterizes all the interactions satisfying the formula $\phi$, and the corresponding operation is $\Delta_{\phi}$.

The restriction term (\ref{eq:restriction-rule}) can be understood as a useful macro-notation for a more complex coordination term forbidding all interactions that involve more than $n$ component instances of type $b$ interacting through port $p$.
If no port $p$ is provided, then the restriction applies to every port of the component type $b$.

Migration terms are built from the given basic rules where operations $\Delta_{p},\Delta_{\phi}$ involve only migration operations.

Coordination terms are built from the given basic rules where operations $\Delta_{p},\Delta_{\phi}$ involve only assignment and reconfiguration operations.
Since coordination terms are defined within the scope of a single motif, the reference to the motif $m$ itself can be omitted.

\subsection{Applications and benchmarks}
\label{subsubsec:benchmarks}
We will now present how some simple application scenarios can be modeled using the \LANG{} coordination language.
For validation purposes and to show possible venues of analysis, the following examples have also been implemented and tested using the \LANG{} Java API.
\begin{example}[Master-Slaves]
	\label{ex:masterslaves-2}
	Let us revisit the scenario of Example \ref{ex:masterslaves-1} using the \LANG{} coordination language. 
	The first step is to generalize the components introduced in Example \ref{ex:masterslaves-0} to \LANG{} component types $Master$ and $Slave$ (Figure~\ref{fig:masterslaves-2}).
	In this case, component types must also provide appropriate local variables that will be used to store the instances to which they get connected to perform the task (i.e. the set of integers $slaves$ for the $Master$ type and the integer $master$ for the $Slaves$ type).
	To restore these local variables to their initial value, we can associate operations $\delta_m = \ASSIGN{}{slaves}{\emptyset}$ and $\delta_s = \ASSIGN{}{master}{0}$ respectively with ports $work$ and $serve$.
	
	The system only requires the definition of a single motif with a trivial map characterized by a single node.
	The coordination term characterizing the desired interaction pattern can be expressed, for instance, using the conjunctive style as follows:
	\begin{align*}
	\rho = \ &\AtMost{1}{Master} \ \&\\
			&\AtMost{1}{Slave.bind} \ \&\\
			&\AtMost{2}{Slave.serve} \ \&\\
			&\irClRULE{\forall m:Master, \exists s:Slave}{\\
			&\quad m.link}{\\
			&\quad \left\| m.slaves \right\| < 2 \wedge s.bind}{\\
			&\quad \quad\ASSIGN{m.}{slaves}{m.slaves \cup \left\lbrace s \right\rbrace}\\
		&} \&\\
		&\irClRULE{\forall s:Slave, m:Master}{\\
			&\quad s.bind}{\\
			&\quad m.link}{\\
			&\quad \ASSIGN{s.}{master}{m}\\
		&} \&\\
	 &\irClRULE{\forall m:Master, \exists s_1:Slave, \exists s_2:Slave}{\\
			&\quad m.work}{\\
			&\quad s_1 \neq s_2 \wedge \left\| m.slaves \right\| = 2 \wedge s_1 \in m.slaves \wedge s_2 \in m.slaves \wedge\\
			&\quad \quad s_1.serve \wedge s_2.serve}{\\
			&\quad \ASSIGN{m.}{buffer}{s_1.mem + s_2.mem}\\
		&} \&
	\end{align*}
	\begin{align*}
	 &\irClRULE{\forall s:Slave, \exists m:Master}{\\
			&\quad s.serve}{\\
			&\quad s.master = m \wedge m.work}{\\
			&\quad \emptyset\\
		&}		
	\end{align*}
	The system model composed by the motif $m$ characterized by $\rho$ and the component types $\left\lbrace Master, Slave \right\rbrace$ can then be initiated with an arbitrary number of component instances of the available types assigned to $m$.
	The resulting system will evolve through interactions that conform to the original description of Example~\ref{ex:masterslaves-1}, meaning that each component instance of type $Master$ will connect with two different component instances of type $Slaves$ (uniquely bound to that same instance of $Master$) and then they will synchronize to carry out the computation.
	Notice that the restriction rules guarantee that only one $Slave$ instance at a time can connect to a single $Master$ instance, and that no more than two $Slave$ instances can participate in an interaction with port $serve$.
	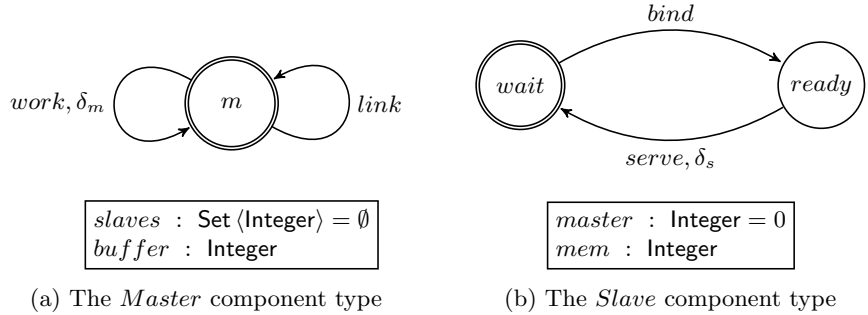
\begin{figure}[h]
		\hspace{0.3cm}
		\begin{subfigure}[b]{0.5\textwidth}
			\centering
			\begin{tikzpicture}[->,>=stealth',shorten >=1pt,auto,node distance=2.8cm,
			semithick]
			\tikzstyle{every state}=[fill=white,text=black]
			
			\node[state,inner sep=9pt,double] (A) at (0,0) {$m$};
			
			\path	(A) edge [loop,min distance=45pt,out=-30,in=30]	node[right] {$link$}		(A)
			(A) edge [loop,min distance=45pt,out=150,in=210]	node[left] {$work,\delta_m$}		(A);
			
			% local variables
			\node[rectangle,draw,align=left] (k1) at (0,-1.75) {
				$\VarDef{slaves}{Set\left\langle Integer \right\rangle} = \emptyset$\\
				$\VarDef{buffer}{Integer}$
			};
			\end{tikzpicture}
			\caption{The $Master$ component type}
		\end{subfigure}
		\begin{subfigure}[b]{0.5\textwidth}
			\centering
			\begin{tikzpicture}[->,>=stealth',shorten >=1pt,auto,node distance=2.8cm,
			semithick]
			\tikzstyle{every state}=[fill=white,text=black]
			
			\node[state,inner sep=5pt,double] (A) at (0,0) {$wait$};
			\node[state] (B) at (4,0) {$ready$};
			
			\path	(A) edge [bend left]	node {$bind$}		(B)
			(B) edge [bend left]	node {$serve,\delta_s$}		(A);
			
			% local variables
			\node[rectangle,draw,align=left] (k1) at (2,-2) {
				$\VarDef{master}{Integer} = 0$\\
				$\VarDef{mem}{Integer}$
			};
			\end{tikzpicture}
			\caption{The $Slave$ component type}
		\end{subfigure}
		\caption{$Master$ and $Slave$ component types}
		\label{fig:masterslaves-2}
	\end{figure}
\end{example}
	We used the \LANG{} Java API to implement the system described in Example~\ref{ex:masterslaves-2} to study the performance of the execution engine when varying the number of component instances in the system.
	For this test we limited the number of execution cycles performed to $20$, and we measured the runtime for systems characterized by $1$ to $8$ $Masters$ and, respectively, $2$ to $16$ $Slaves$.
	
\begin{figure}[h]
		\centering
		\includegraphics[width=0.8\linewidth]{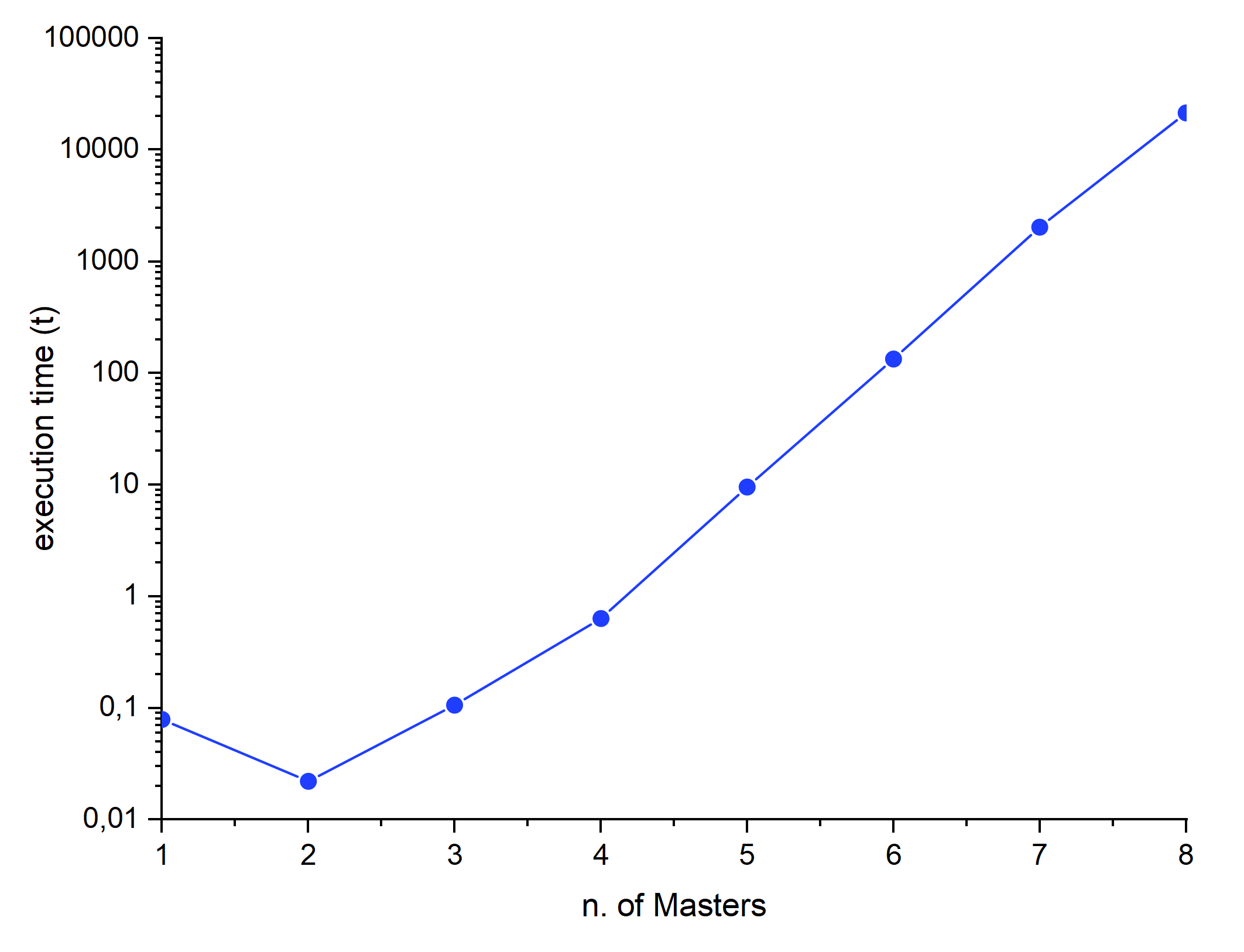}
		\caption{Runtime of $20$ execution cycles of the implementation of Example~\ref{ex:masterslaves-2}}
		\label{fig:graph-masterslaves}
	\end{figure}

	The results are illustrated in Figure~\ref{fig:graph-masterslaves}.
	The exponential growth in the runtime with the number of components is caused by the fact that the current implementation of the execution engine searches exhaustively over the set of all possible interactions collecting all the maximal ones, and then selects one at random.
	
\begin{example}[Coordinating flocks of interacting robots]
	\label{ex:stigmergy-1}
	Consider a system with $N$ robots moving in a square grid, each one with given initial location and initial movement direction.
	Robots are equipped with a sensor that can detect other peers within a specific range $r$ and assess their direction: when this happens, the robot changes its own direction accordingly.
	
	We require that robots maintain a timestamp of their last interaction with another peer: when two robots are within the range of their sensors, their direction is updated with the one having the highest timestamp.
	For the sake of simplicity we also assume that the grid is, in fact, a torus with no borders.
	
	To model these robots in \LANG{} we will define a $Robot$ component type as the one represented in Figure~{\ref{fig:robot-1}}.
	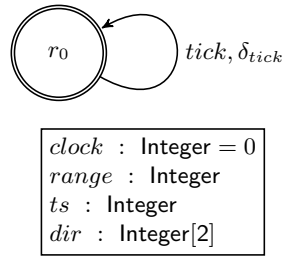
\begin{figure}[h]
		\begin{tikzpicture}[->,>=stealth',shorten >=1pt,auto,node distance=2.8cm,
		semithick]
		\tikzstyle{every state}=[fill=white,text=black]
		
		\node[state,inner sep=9pt,double] (A) at (0,0) {$r_0$};
		
		\path	(A) edge [loop,min distance=45pt,out=-30,in=30]	node[right] {$tick,\delta_{tick}$}		(A);
		
		% local variables
		\node[rectangle,draw,align=left] (k1) at (1.25,-1.85) {
			$\VarDef{clock}{Integer} = 0$\\
			$\VarDef{range}{Integer}$\\
			$\VarDef{ts}{Integer}$\\
			$\VarDef{dir}{Integer[2]}$
		};
		\end{tikzpicture}
		\centering
		\caption{The $Robot$ component type}
		\label{fig:robot-1}
	\end{figure}
	Each robot maintains a local $clock$ that is incremented by 1 through an assignment statement in $\delta_{tick}$ every time an instance interacts with port $tick$.
	
	A motif that realizes the described scenario can be defined with the conjunction of two rules: one that enforces synchronization between every $Robot$ instance through port $tick$ allowing information exchange when possible, and another that enables all robots performing a $tick$ to move:
	\begin{align*}
	\rho = \ &\irClRULE{\forall r:Robot}{r.tick}{\TRUE}{\MOVE{r}{\MVAt(r)+r.dir}}\\
	& \&\\
	 &\irClRULE{\forall r_1,r_2:Robot}{r_1.tick}{r_2.tick}{\\
		&\quad \IFTHEN{r_1 \neq r_2}{\\
		&\quad \quad \IFTHEN{distance(\MVAt(r_1),\MVAt(r_2)) < r_1.range \ \wedge \\
			&\quad \quad \quad \quad \left( r_1.ts < r_2.ts \vee \left( r_1.ts = r_2.ts \wedge r_1 < r_2 \right) \right)}{\\
		&\quad \quad \quad \ASSIGN{r_1.}{dir}{r_2.dir} ; \ \ASSIGN{r_1.}{ts}{r_1.clock} ; \ \ASSIGN{r_2.}{ts}{r_2.clock}}}} 
	\end{align*}
	where:
	\begin{itemize}
		\item we use a map whose nodes are addressed via size-two integer arrays $\left[x,y\right]$;
		\item we are using a $distance(n_1,n_2)$ function that returns the euclidean distance between two points in an n-dimensional space;
		\item in the inequality $r_1 < r_2$ we use instance variables $r_1,r_2$ in place of their respective integer instance identifiers.
	\end{itemize}
	The rule $\rho$, which adopts the conjunctive style, can be intuitively understood breaking it into two parts:
	\begin{enumerate}
		\item every robot $r$ can interact with its port $r.tick$, and if it does it also moves according to its stored direction $r.dir$\footnote{Notice that if the direction of a robot is updated at a given time, the robot will move according to this new direction only during the next clock cycle because of the adopted snapshot semantics.};
		\item for every robot $r_1$ to interact with its port $r_1.tick$, every robot $r_2$ must also participate in interaction with its port $r_2.tick$ (i.e. interactions through port $tick$ are strictly synchronous). 
		Furthermore, for every pair of distinct ($r_1 \neq r_2$) robots $r_1,r_2$ interacting through their respective $tick$ ports: if they are closer than a given range ($r_1.range$) and either $r_1$ has updated its direction less recently ($r_1.ts < r_2.ts$) or they have updated their directions at the same time but $r_2$ has a higher instance identifier ($r_1.ts = r_2.ts \wedge r_1 < r_2$)\footnote{Since all robots synchronize on the same ``clock'', many of them might update their respective directions differently at the same time: adding the ``tiebreaker'' on the instance identifier when timestamps are equal allows data exchange even in these cases.}, then $r_1$ will update its direction and timestamp using $r_2$'s.
	\end{enumerate}
	
\end{example}

	We used the \LANG{} Java API to implement the system described in Example~\ref{ex:stigmergy-1} and study its behaviour while varying the size of the grid and the communication range for a fixed number of robots.
	Intuitively, we expect to observe a faster convergence in the movement directions as the size of the grid shrinks and/or as the communication range increases.
	
	We fix the number of robots in the system to $9$, and we choose a specific initial direction for each one of them.
	We also choose the same $range$ value for all robots.
	The mapping of the robots to a grid of size $s \times s$ is realized in such a way that they are uniformly spaced both horizontally and vertically. 
	We choose grid sizes proportional to $3$ for uniformity.
	
\begin{figure}[h]
		\centering
		\includegraphics[width=0.8\linewidth]{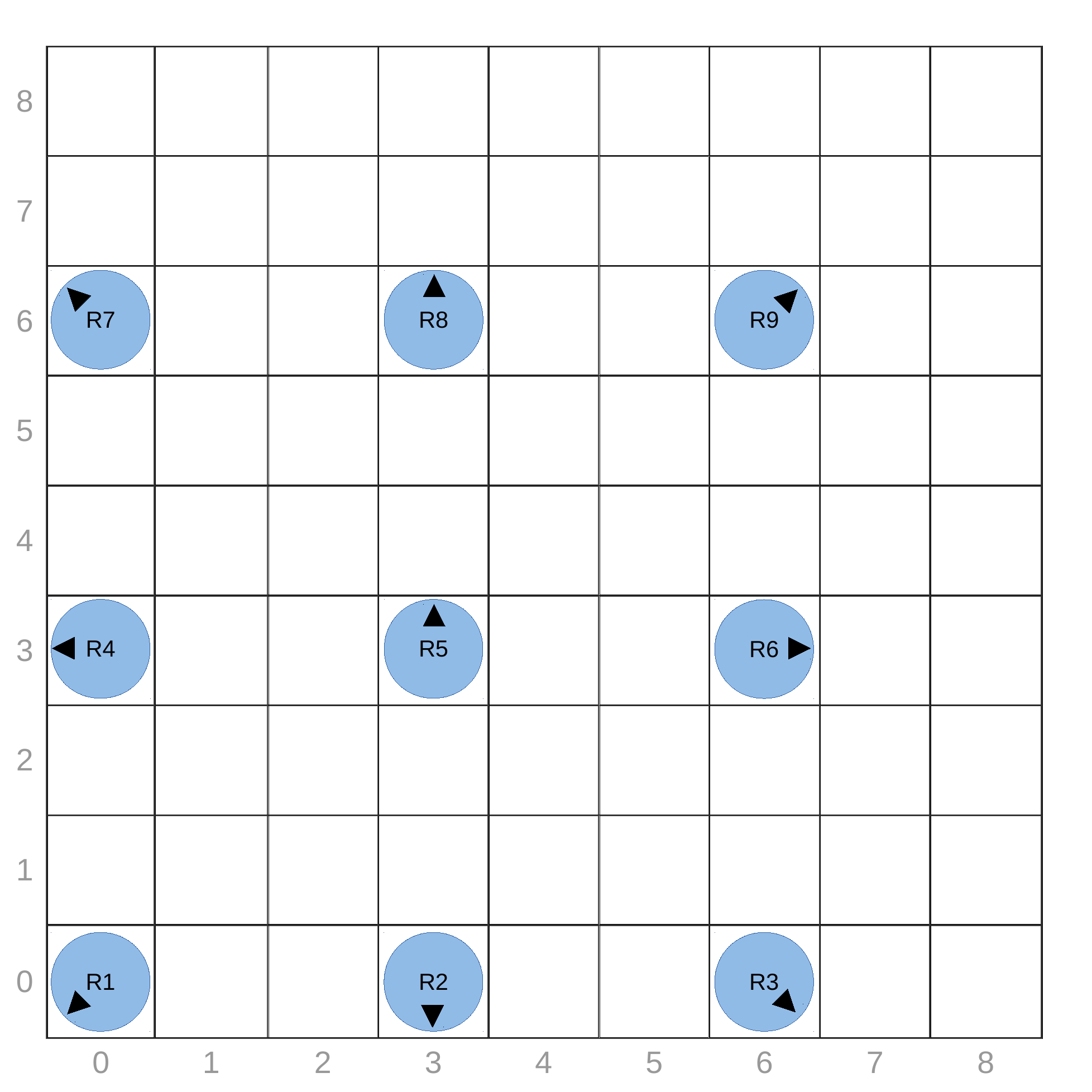}
		\caption{Initial system configuration for grid size $s=9$}
		\label{fig:robots-grid}
	\end{figure}

	An example of the initial configuration for a grid of size $s=9$ is shown in Figure~\ref{fig:robots-grid}.

\begin{figure}[h]
		\centering
		\begin{subfigure}[b]{0.49\linewidth}
			\centering
			\includegraphics[width=1\linewidth]{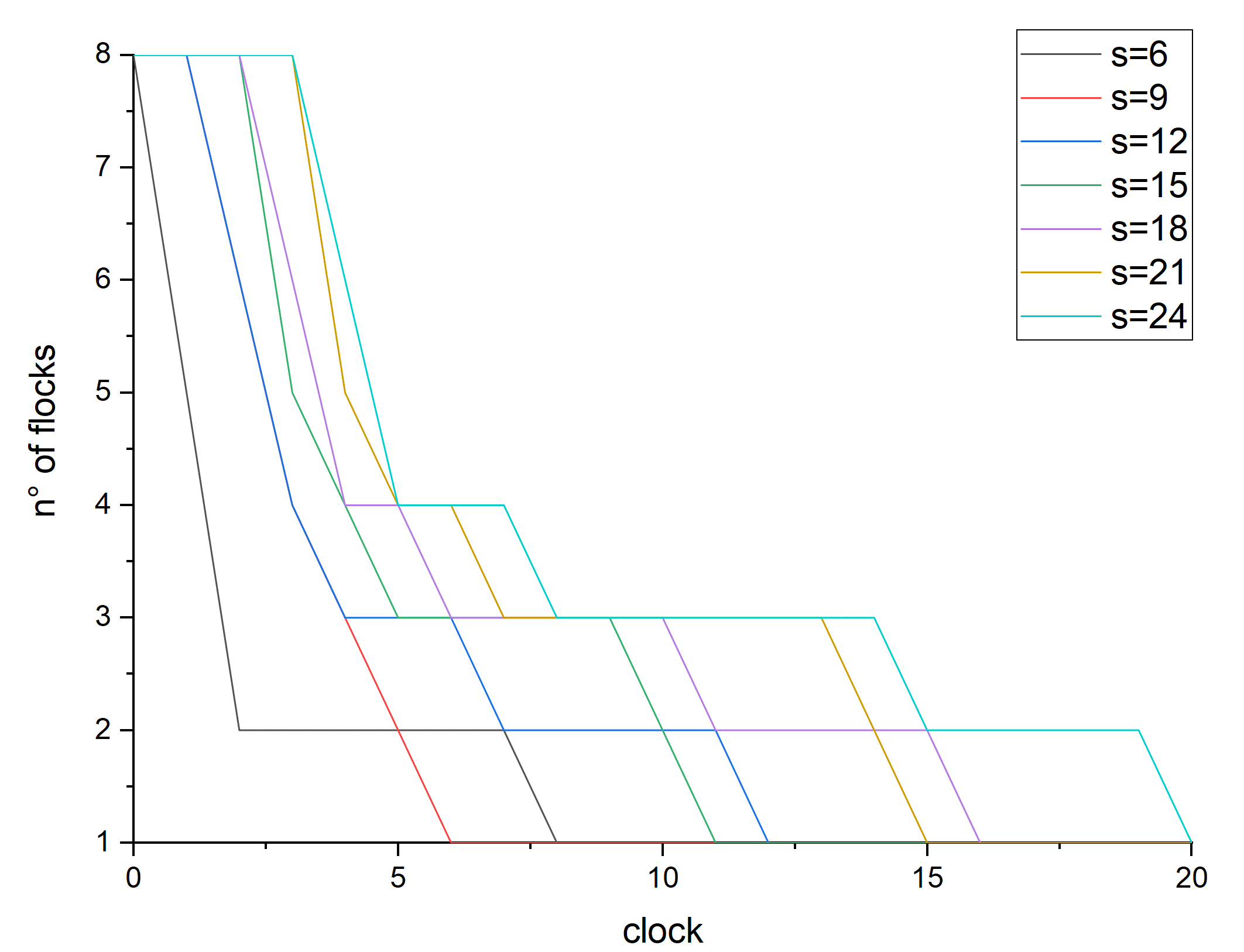}
			\caption{$range=3$}
			\label{subfig:commrobot-3}
		\end{subfigure}
		\begin{subfigure}[b]{0.49\linewidth}
			\centering
			\includegraphics[width=1\linewidth]{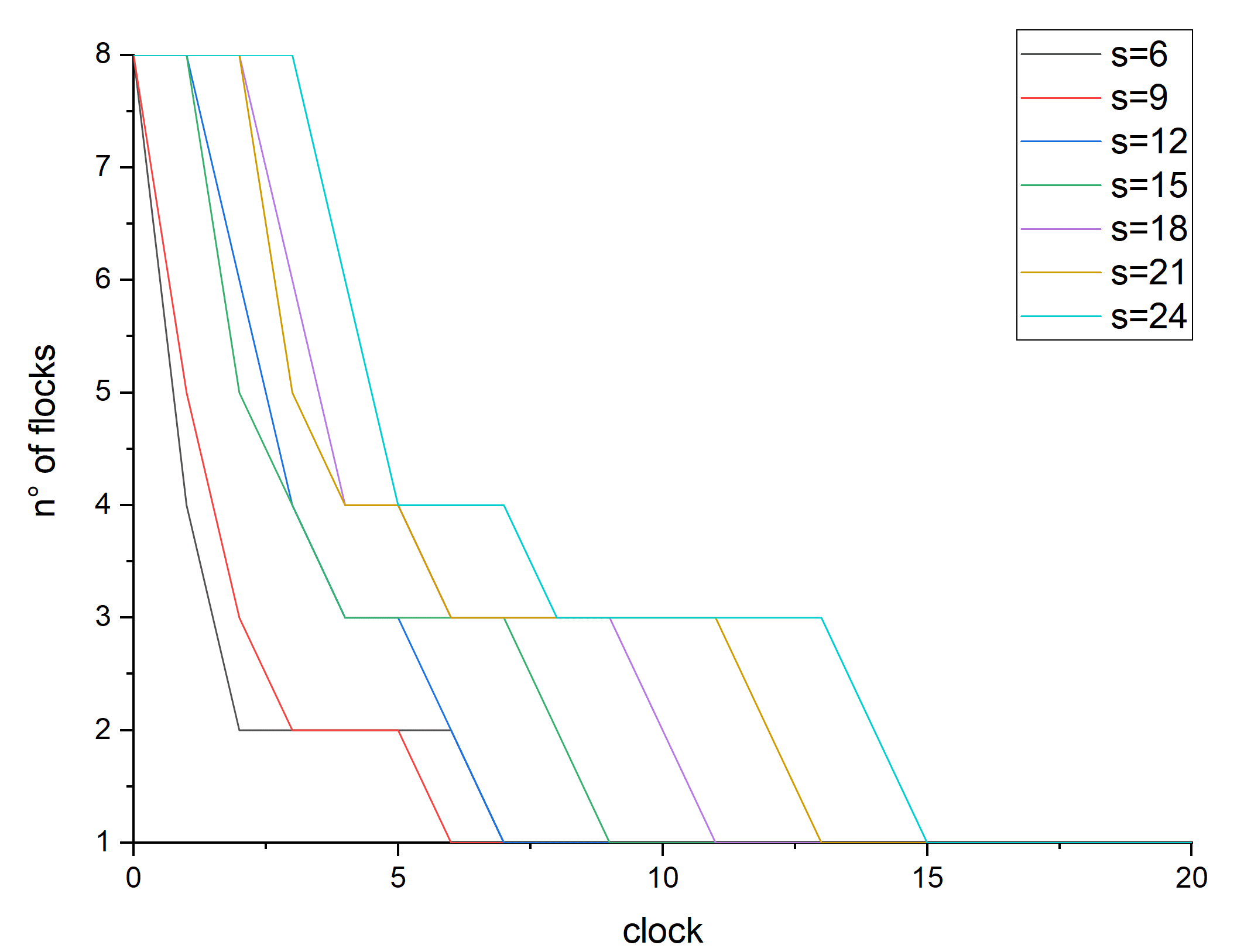}
			\caption{$range=4$}
			\label{subfig:commrobot-4}
		\end{subfigure}
		\begin{subfigure}[b]{0.49\linewidth}
			\centering
			\includegraphics[width=1\linewidth]{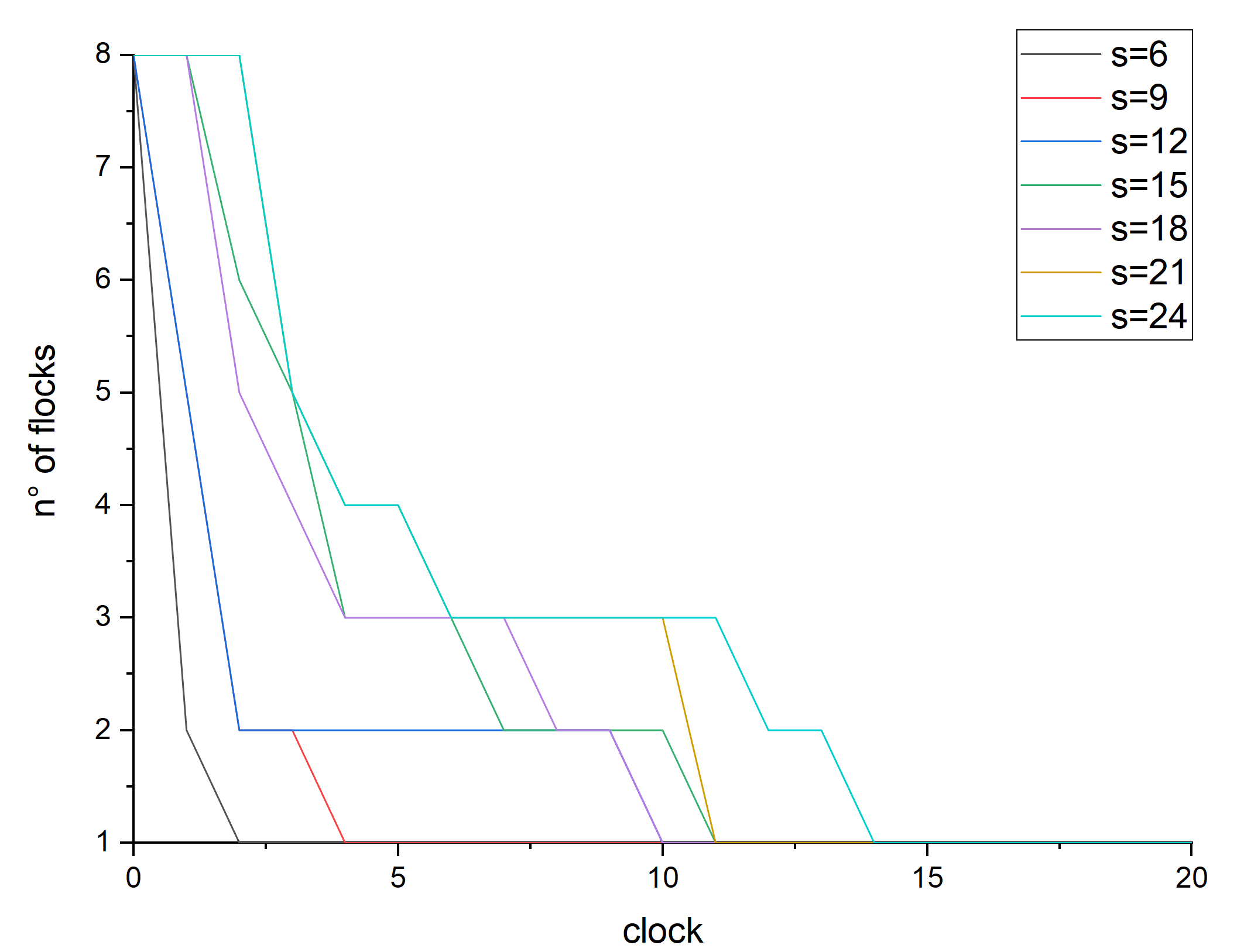}
			\caption{$range=5$}
			\label{subfig:commrobot-5}
		\end{subfigure}
		\begin{subfigure}[b]{0.49\linewidth}
			\centering
			\includegraphics[width=1\linewidth]{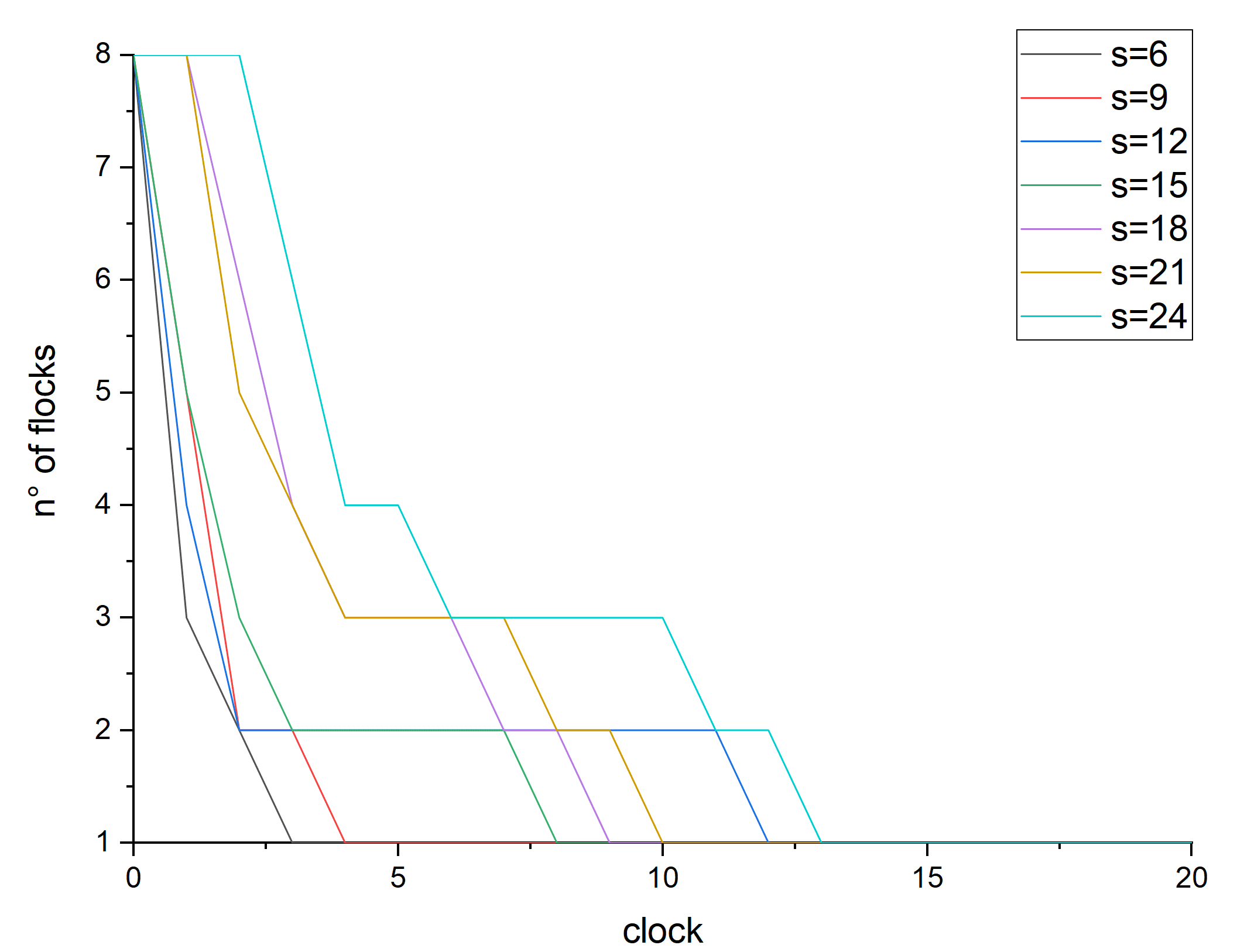}
			\caption{$range=6$}
			\label{subfig:commrobot-6}
		\end{subfigure}
		\caption{Trends in the number of flocks over time at different communication ranges}
		\label{fig:graph-communicatingrobots}
	\end{figure}

	The graphs in Figure~\ref{fig:graph-communicatingrobots} show the trend in the number of flocks (i.e., the number of groups of robots moving according to a different direction) over time for different values of the given $range$ of communication.

	Indeed, the results confirm our expectations: the adopted initial setup procedure of the robot's positions and directions allows them to converge to an homogeneous flock within 20 clock ticks, a number which decreases as we increase the communication range.
	There is also an opposite trend when increasing the size of the grid, although it is interesting to see that there are several exceptions to this rule (e.g. for $range=3$ convergence on the grid $s=6$ takes more time than on the grid $s=9$; the same applies for $s=12$ vs $s=15$ and $s=18$ vs $s=21$).
	
%	Furthermore, we can see that, for the given initial configuration, with $range=3$ and $range=4$ convergence cannot be achieved when the grid size exceeds a threshold\footnote{Non-convergence is not related to a limited observation over time: the graphs in Figure~\ref{fig:graph-communicatingrobots} only show the significant portion of the measured data that ranges over 500 clock ticks. In these cases the resulting flocks end up moving over parallel paths, but according to opposite directions.}.

\begin{example}[Coordinating flocks of robots with stigmergy]
	\label{ex:stigmergy-2}
	We consider a variant of the previous problem by using stigmergy \cite{pinciroli2015buzz}.
%	This coordination mechanism relies on a more decoupled form of data exchange that does not involve direct interaction between agents.
%	Instead, agents interact with the environment, that acts as a shared memory that can be queried and modified.
%	
%	We consider the same scenario of Example \ref{ex:stigmergy-1}, but instead of letting robots sense each other we will allow them to ``mark'' their locations with their direction and an associated timestamp.
%	In this way, each time a robot moves to a node in the map it will either update its direction with the one stored in the node or update the one associated with the node with the direction of the robot (depending on whether the timestamp is higher than the last time the robot changed its direction or not).

	Instead of letting robots sense each other, we will allow them to ``mark'' their locations with their direction and an associated timestamp. 
	In this way, each time a robot moves to a node in the map it will either	update its direction with the one stored in the node or update the one associated with the node with the direction of the robot (depending on whether the timestamp is higher than the last time the robot changed its direction or not).
	
	The $Robot$ component type represented in Figure~\ref{fig:robot-1} can still be used without modifications (the $range$ local variable will be ignored).	
		
	The rule associated to the motif becomes:
	\begin{align*}
	\rho' = \ &\rcClRULE{\forall r:Robot}{\\
		&\quad r.tick}{\\
		&\quad \IFTHENELSE{\MVAt(r).ts > r.ts}{\\
			&\quad \quad \ASSIGN{r.}{dir}{\MVAt(r).dir} ; \ \ASSIGN{r.}{ts}{r.clock} ; \ \ASSIGN{\MVAt(r).}{ts}{r.clock}
			\\& \quad}{\\
			&\quad \quad \ASSIGN{\MVAt(r).}{ts}{r.clock} ; \ \ASSIGN{\MVAt(r).}{dir}{r.dir}}\\
		&\quad \MOVE{r}{\MVAt(r)+r.dir}}
	\end{align*}
	Notice that we are now adopting a disjunctive-style specification for $\rho'$.
	We can interpret the rule $\rho'$ as follows:
	\begin{enumerate}
		\item every robot $r$ must participate in all interactions with its port $r.tick$, and will move in the map according to its stored direction $r.dir$;
		\item every robot $r$ either updates its direction with the one stored in the node $\MVAt(r)$ if the latter is more recent (i.e., if $\MVAt(r).ts > r.ts$) or overwrites the direction stored in the node with its own otherwise.
	\end{enumerate}

\end{example}

Example~\ref{ex:stigmergy-2} has also been implemented using the \LANG{} Java API.
%In order for the results of the two examples on coordination of robots to be comparable, we will fix the same parameters as in Example~\ref{ex:stigmergy-1} regarding number of robots, initial directions, set of tested grid sizes and mapping criterion.
For a comparison with Example~\ref{ex:stigmergy-1}, we fix the same parameters as in regarding number of robots, initial directions, set of tested grid sizes and mapping criterion.

\begin{figure}[h]
	\centering
	\includegraphics[width=0.8\linewidth]{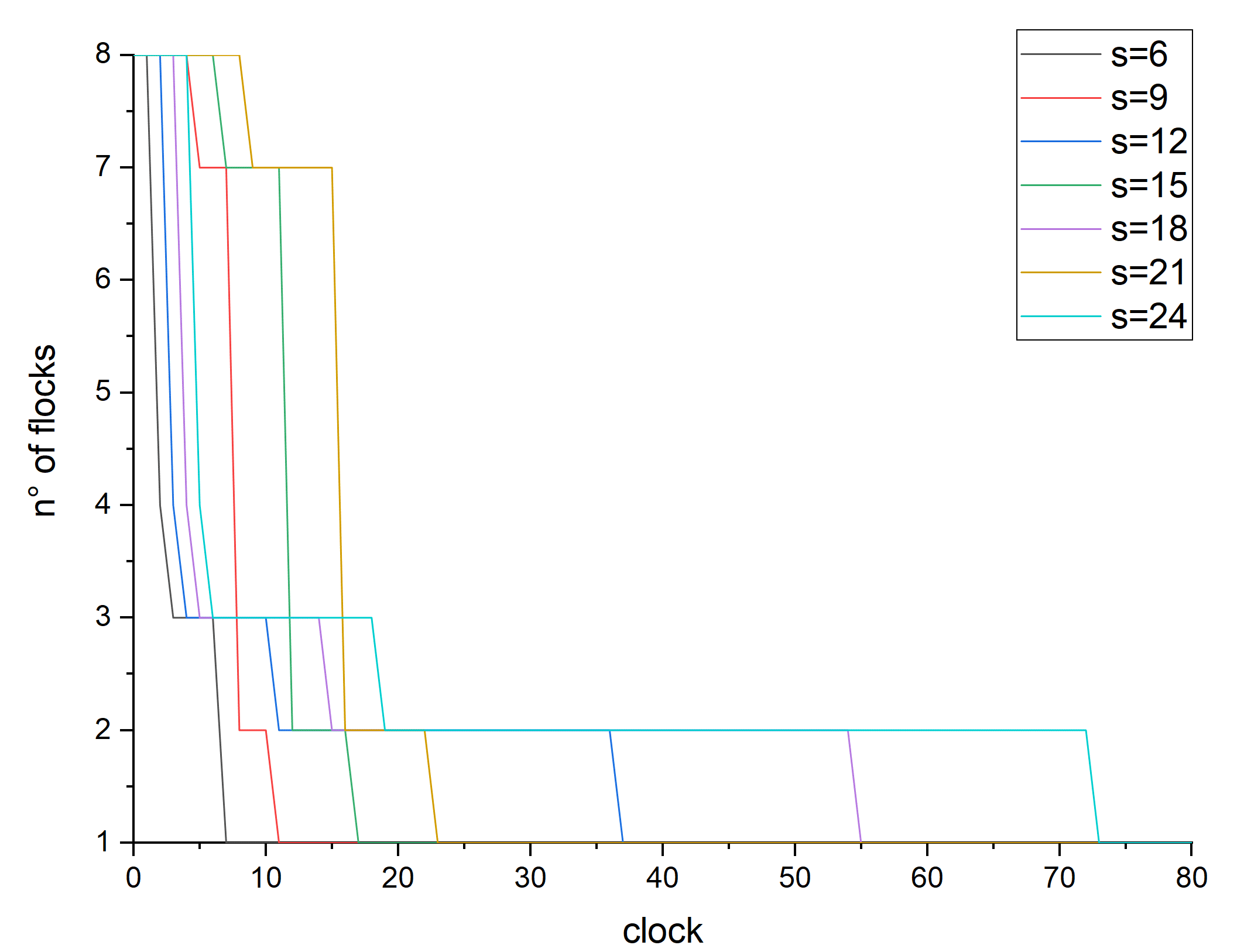}
	\caption{Evolution of the number of flocks over time at different communication ranges}
	\label{fig:graph-stigmergicrobots}
\end{figure}

We can reasonably expect a similar correlation between convergence time and grid size as in the case for communicating robots. 
Indeed, this is confirmed by the graph in Figure~\ref{fig:graph-stigmergicrobots}, which shows the trends in the number of flocks for different grid sizes.

It is worth observing that convergence time and grid size are, again, not always directly proportional: here it is particularly striking how the robots converge to a single flock for grid sizes equal to $15$ and $21$ in roughly half the time it takes for them to converge on the smaller grid with $s=12$.

\begin{figure}[h]
	\centering
	\begin{subfigure}[b]{0.49\linewidth}
		\centering
		\includegraphics[width=1\linewidth]{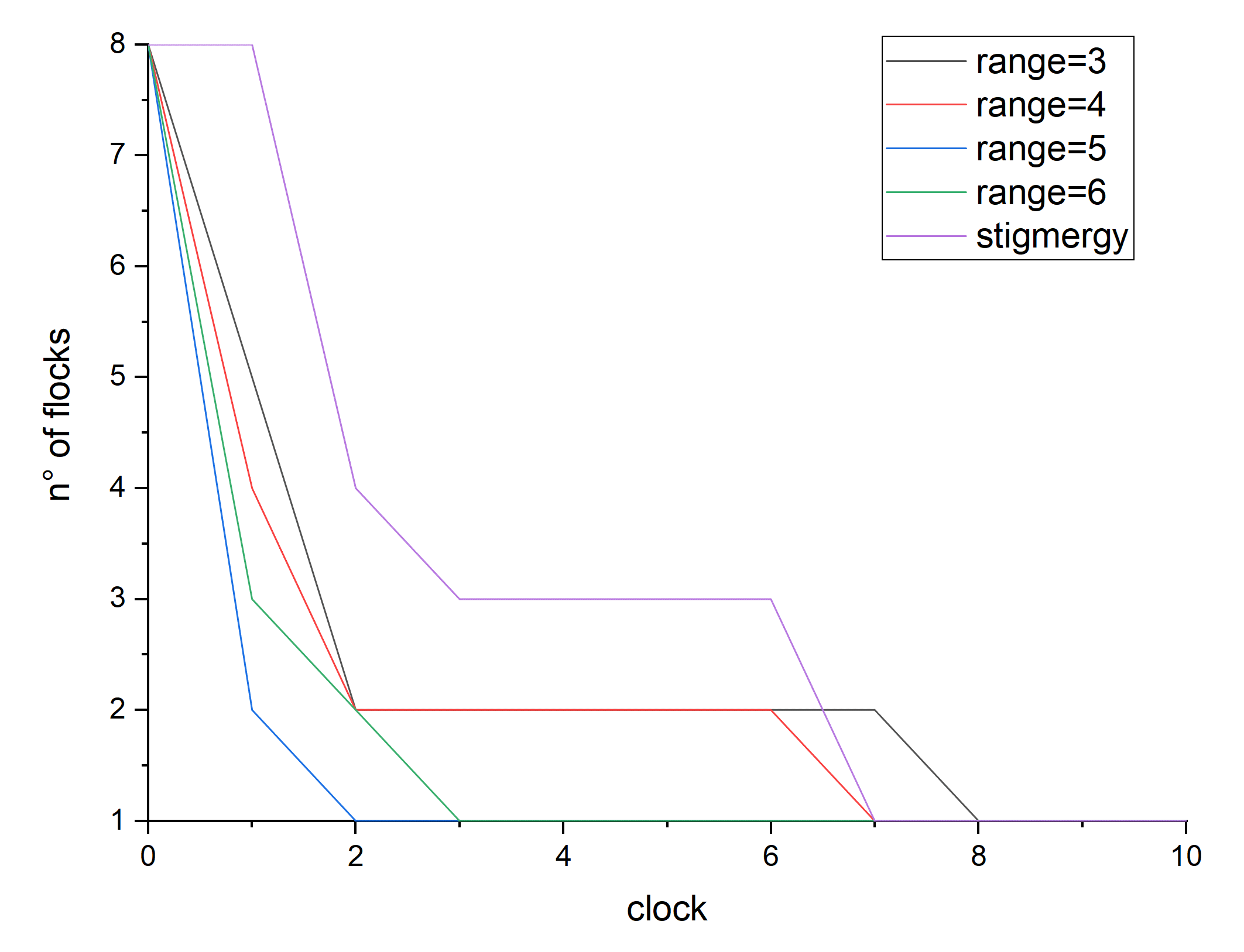}
		\caption{$s=6$}
		\label{subfig:robots-compare6}
	\end{subfigure}
	\begin{subfigure}[b]{0.49\linewidth}
		\centering
		\includegraphics[width=1\linewidth]{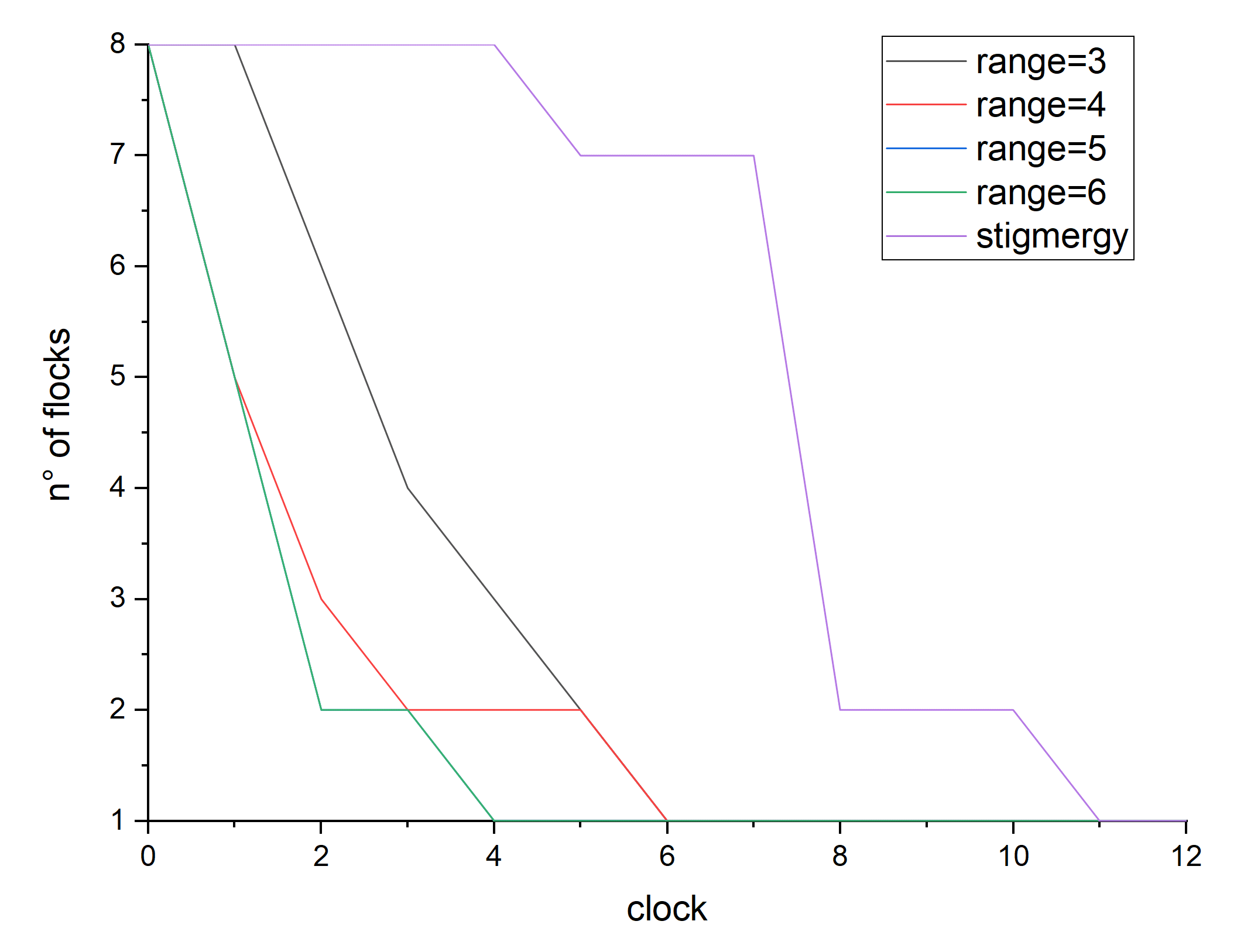}
		\caption{$s=9$}
		\label{subfig:robots-compare9}
	\end{subfigure}
	\begin{subfigure}[b]{0.49\linewidth}
		\centering
		\includegraphics[width=1\linewidth]{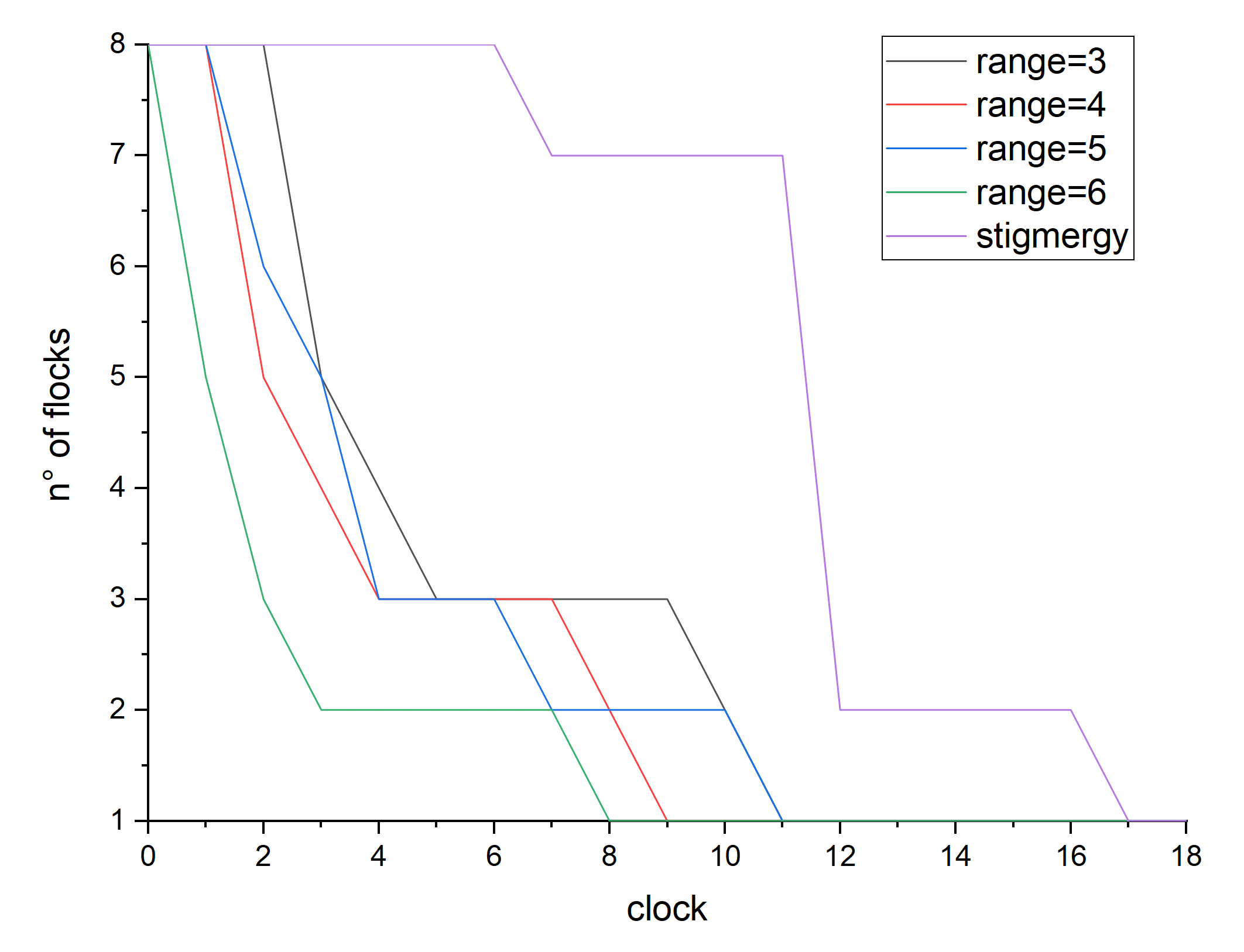}
		\caption{$s=15$}
		\label{subfig:robots-compare15}
	\end{subfigure}
	\begin{subfigure}[b]{0.49\linewidth}
		\centering
		\includegraphics[width=1\linewidth]{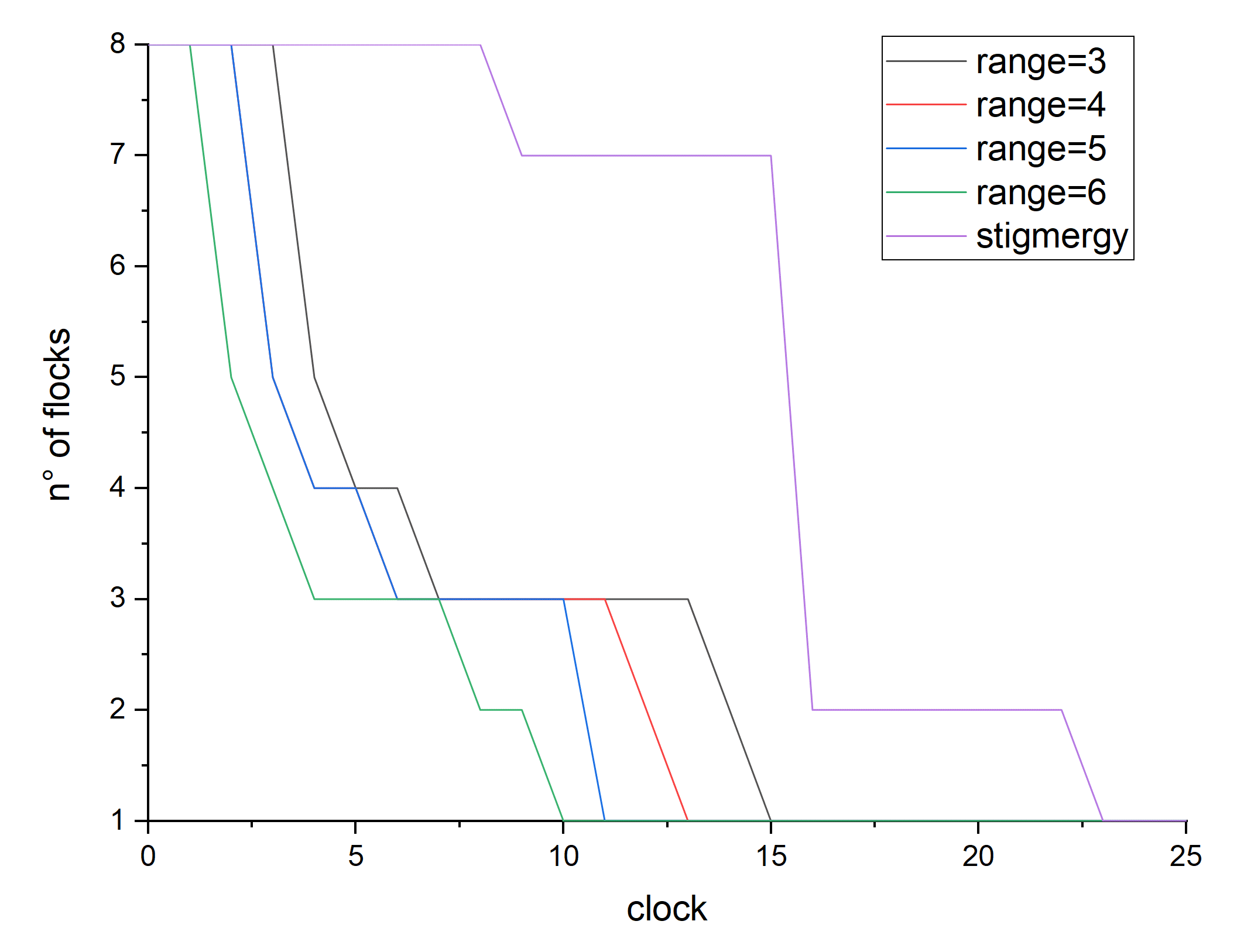}
		\caption{$s=21$}
		\label{subfig:robots-compare21}
	\end{subfigure}
	\caption{Comparison between the two approaches at different grid sizes}
	\label{fig:graph-comparison}
\end{figure}

The graphs in Figure~\ref{fig:graph-comparison} compare directly the convergence trends using the two approaches on grids of different sizes.
From these we can appreciate how the stigmergy-based solution performs roughly on-par with the interaction-based one for small maps, progressively losing ground to the latter as the map becomes larger.
This comparison also helps to better visualize how the implementation not resorting on sensors initially requires some time to populate the map with information which is proportional with the size of the map itself.

	% !TEX root =../DreAM-article.tex

\section{Related work}
\label{sec:relatedwork}

\LANG{} allows both conjunctive and disjunctive style modeling of dynamic reconfigurable systems. 
It inherits the expressiveness of the coordination mechanisms of BIP \cite{bliudze2008algebra} as it directly encompasses multiparty interaction and extends previous work on modeling parametric architectures \cite{bozga2012modeling} in many respects. 
In \LANG{} interactions involve not only transfer of values but also encompass reconfiguration and self-organization by relying on the notions of maps and motifs.

When the disjunctive style is adopted, \LANG{} can be considered as an exogenous coordination language, e.g., an ADL. 
A comparison with the many ADL's is beyond the scope of the paper.
Nonetheless, to the best of our knowledge \LANG{} surpasses existing exogenous coordination frameworks in that it offers a well-thought and methodologically complete set of primitives and concepts.

When conjunctive style is adopted, \LANG{} can be used as an endogenous coordination language comparable to process calculi to the extent they rely on a single associative parallel composition operator. 
In \LANG{} this operator is logical conjunction. 
It is easy to show that for existing process calculi parallel composition is a specialization of conjunction in Interaction Logic. 
For CCS \cite{milner1980calculus} the causal rules are of the form $p \Rightarrow \bar{p}$, where $p$ and $\bar{p}$ are input and output port names corresponding to port symbol $p$. 
For CSP \cite{brookes1984theory}, the causal rules implementing the interface parallel operator parameterized by the channel $a$ are of the form $a_i \Rightarrow \bigwedge_{a_j \in A}a_j$, where $A$ is the set of ports communicating through $a$. 

Also other richer calculi, such as $\pi$-calculus \cite{milner1992calculus}, that offer the possibility of modeling dynamic infrastructure via channel passing can be modeled in \LANG{} with its reconfiguration operations.
%\warn{Give an example?}
% \warn{Here say something more specific about the dynamic features of process calculi. Clearly parametricity is not an issue can be easily modeled using arrays of process variables.  A short example could illustrate the differences}
%
Formalisms with richer communication models, such as AbC \cite{AbCFORTE}, offering multicasting communications by selecting groups of partners according to predicates over their attributes, can also be rendered in  \LANG{}.
Attribute based interaction can be simulated by our interaction mechanism involving guards on the exchanged values and atomic transfer of values. 
%\warn{Give an example?}

\LANG{}  was designed with autonomy in mind. As such it has some similarities with languages for autonomous systems in particular robotic systems such as Buzz \cite{pinciroli2015buzz,pinciroli2016buzz}. Nonetheless, our framework is more general as it does not adopt assumptions about timed synchronous cyclic behavior of components.
%Similarities include focus on the collective behavior of agents organized in swarms and sharing a data structure for stigmergy that turns out to be our map. 

The relationships between our approach and graph based architectural description languages such as ADR\cite{bruni2007style} and HDR\cite{bruni2009hierarchical} will be the subject of future work.

Finally, \LANG{} shares the same conceptual framework with DR-BIP\cite{drbip}. 
The latter is an extension of BIP with component dynamism and reconfiguration. 
As such it adopts an exogenous and imperative approach based on the use of connectors. 
A detailed comparison between \LANG{} and DR-BIP will be the object of a forthcoming publication.
	
	\section{Discussion}
\label{sec:discussion}

We have proposed a framework for the description of dynamic reconfigurable systems supporting their incremental construction according to a hierarchy of structuring concepts going from components to sets of motifs forming a system. 
Such a hierarchy guarantees enhanced expressiveness and incremental modifiability thanks to the following features:

{\bf Incremental modifiability of models at all levels}:
The interaction rules associated with a component in a motif can be modified and composed independently. 
Components can be defined independently of the maps and their context of use in a motif. 
Self-organization can be modeled by combining motifs, i.e., system modes for which particular interaction rules hold.

{\bf Expressiveness}: This is inherited from BIP as the possibility to directly specify any kind of static coordination without modifying the involved components or adding extra coordinating components. 
Regarding dynamic coordination, the proposed language directly encompasses the identified levels of dynamicity by supporting component types and the expressive power of first order logic. 
Nonetheless, explicit handling of quantifiers is limited to declarations that link component names to coordinates. 

{\bf Flexible Semantics}: The language relies on an operational semantics that admits a variety of implementations between two extreme cases. 
One consists in precomputing a global interaction constraint applied to an unstructured set of component instances and choosing the enabled interactions and the corresponding operations for a given configuration. 
The other consists in computing separately interactions for motifs or groups and combining them. 

The results about the relationship between conjunctive and disjunctive styles show that while they are both equally expressive for interactions without data transfer, the disjunctive style is more expressive when interactions involve data transfer. 
We plan to further investigate this relationship to characterize more precisely this limitation that seems to be inherent to modular specification.

All results are too recent and  many open avenues need to be explored.
The language and its tools should be evaluated against real-life mobile applications 
such as autonomous transport systems, swarm robotics or  telecommunication systems.

	\bibliographystyle{ieeetr}
	\bibliography{DReAM-article}

\begin{thebibliography}{10}

\bibitem{garlan2014software}
D.~Garlan, ``Software architecture: a travelogue,'' in {\em Proceedings of the
  on Future of Software Engineering}, pp.~29--39, ACM, 2014.

\bibitem{taivalsaari2014liquid}
A.~Taivalsaari, T.~Mikkonen, and K.~Syst{\"a}, ``Liquid software manifesto: the
  era of multiple device ownership and its implications for software
  architecture,'' in {\em Proc. 38th Computer Software and Applications
  Conference}, pp.~338--343, IEEE, 2014.

\bibitem{bradbury2004organizing}
J.~S. Bradbury, ``Organizing definitions and formalisms for dynamic software
  architectures,'' {\em Technical Report}, vol.~477, 2004.

\bibitem{oreizy1998issues}
P.~Oreizy {\em et~al.}, ``Issues in modeling and analyzing dynamic software
  architectures,'' in {\em Proc. Int'l Workshop on the Role of Software
  Architecture in Testing and Analysis}, pp.~54--57, 1998.

\bibitem{malavolta2013industry}
I.~Malavolta, P.~Lago, H.~Muccini, P.~Pelliccione, and A.~Tang, ``What industry
  needs from architectural languages: A survey,'' {\em IEEE Transactions on
  Software Engineering}, vol.~39, no.~6, pp.~869--891, 2013.

\bibitem{butting2017classification}
A.~Butting, R.~Heim, O.~Kautz, J.~O. Ringert, B.~Rumpe, and A.~Wortmann, ``A
  classification of dynamic reconfiguration in component and connector
  architecture description languages,'' in {\em Pre-proc. 4th Int'l Workshop on
  Interplay of Model-Driven and Component-Based Software Engineering}, p.~13,
  2017.

\bibitem{medvidovic2007moving}
N.~Medvidovic, E.~M. Dashofy, and R.~N. Taylor, ``Moving architectural
  description from under the technology lamppost,'' {\em Information and
  Software Technology}, vol.~49, no.~1, pp.~12--31, 2007.

\bibitem{bliudze2008algebra}
S.~Bliudze and J.~Sifakis, ``The algebra of connectors - structuring
  interaction in {BIP},'' {\em IEEE Transactions on Computers}, vol.~57,
  no.~10, pp.~1315--1330, 2008.

\bibitem{pinciroli2015buzz}
C.~Pinciroli, A.~Lee-Brown, and G.~Beltrame, ``Buzz: An extensible programming
  language for self-organizing heterogeneous robot swarms,'' {\em arXiv
  preprint arXiv:1507.05946}, 2015.

\bibitem{bozga2012modeling}
M.~Bozga, M.~Jaber, N.~Maris, and J.~Sifakis, ``Modeling dynamic architectures
  using {Dy-BIP},'' in {\em Software Composition}, pp.~1--16, Springer, 2012.

\bibitem{milner1980calculus}
R.~Milner, ``A calculus of communicating systems,'' 1980.

\bibitem{brookes1984theory}
S.~D. Brookes, C.~A. Hoare, and A.~W. Roscoe, ``A theory of communicating
  sequential processes,'' {\em Journal of the ACM}, vol.~31, no.~3,
  pp.~560--599, 1984.

\bibitem{milner1992calculus}
R.~Milner, J.~Parrow, and D.~Walker, ``{A Calculus Of Mobile Processes, I},''
  {\em Information and computation}, vol.~100, no.~1, pp.~1--40, 1992.

\bibitem{AbCFORTE}
Y.~{Abd Alrahman}, R.~{De Nicola}, and M.~Loreti, ``On the power of
  attribute-based communication,'' in {\em Proc. Formal Techniques for
  Distributed Objects, Components, and Systems - {FORTE} 2016 - 36th {IFIP}
  {WG} 6.1 In'l Conference}, pp.~1--18, 2016.

\bibitem{pinciroli2016buzz}
C.~Pinciroli and G.~Beltrame, ``Buzz: An extensible programming language for
  heterogeneous swarm robotics,'' in {\em Intelligent Robots and Systems
  (IROS), 2016 IEEE/RSJ International Conference on}, pp.~3794--3800, IEEE,
  2016.

\bibitem{bruni2007style}
R.~Bruni, A.~L. Lafuente, U.~Montanari, and E.~Tuosto, ``Style based
  reconfigurations of software architectures,'' {\em Universita di Pisa, Tech.
  Rep. TR-07-17}, 2007.

\bibitem{bruni2009hierarchical}
R.~Bruni, A.~Lluch-Lafuente, and U.~Montanari, ``Hierarchical design rewriting
  with maude,'' {\em Electronic Notes in Theoretical Computer Science},
  vol.~238, no.~3, pp.~45--62, 2009.

\bibitem{drbip}
R.~El~Ballouli, S.~Bensalem, M.~Bozga, and J.~Sifakis, ``Four exercises in
  programming dynamic reconfigurable systems: methodology and solution in
  {DR-BIP},'' in {\em {ISoLA 2018}}, vol.~11246, Springer, 2018.

\end{thebibliography}
	
%	\newpage
%	\appendix
%	\input{tex/appendix}
\end{document}